\documentstyle[tighten,preprint,amsfonts,eqsecnum,aps,prd]{revtex}
\begin{document}
\draft

\hyphenation{
mani-fold
mani-folds
geo-metry
geo-met-ric
}


\newcommand{\zr}{\mbox{$\frak r$}}
\newcommand{\zp}{\mbox{$\frak p$}}
\newcommand{\zH}{\mbox{$\cal H$}}
\newcommand{\zN}{\mbox{$\cal N$}}
\newcommand{\zP}{\mbox{$\cal P$}}

\def\fallpar{\beta}
\def\ridgepar{\gamma}
\def\shellpar{l}
\def\extder{\bbox{\delta}}
\def\boldGamma{\bbox{\Gamma}}
\def\boldGammahat{\bbox{\hat\Gamma}}
\def\bp{{\bf p}}
\def\bh{{\bf h}}

\def\transsurface{{\bar H}}

\def\sfH{{\sf H}}

\def\BbbR{{\Bbb R}}
\def\RPthree{{{\Bbb RP}^3}}
\def\RPtwo{{{\Bbb RP}^2}}

\def\casehalf{{\case{1}{2}}}


\preprint{\vbox{\baselineskip=12pt
\rightline{PP97--120}
\rightline{WISC--MILW--97--TH--11}
\rightline{gr-qc/9706051}}}
\title{Reduced phase space formalism for
spherically symmetric geometry
with a massive dust shell}
\author{John L. Friedman\footnote{Electronic address:
friedman@thales.phys.uwm.edu}}
\address{
Department of Physics,
University of
Wisconsin--Milwaukee,
\\
P.O.\ Box 413,
Milwaukee, Wisconsin 53201, USA}
\author{Jorma Louko\footnote{On leave of absence from
Department of Physics, University of Helsinki.
Present address:
Max-Planck-Institut f\"ur Gravitations\-physik,
Schlaatzweg~1,
D--14473 Potsdam,
Germany.
Electronic address:
louko@aei-potsdam.mpg.de
}}
\address{
Department of Physics,
University of Maryland,
College Park,
Maryland 20742--4111,
USA}
\author{Stephen N. Winters-Hilt\footnote{Electronic address:
winters@csd.uwm.edu}}
\address{
Department of Physics,
University of
Wisconsin--Milwaukee,
\\
P.O.\ Box 413,
Milwaukee, Wisconsin 53201, USA}
\date{Revised version, October 1997. 
Published in Phys.\ Rev.\ D {\bf 56} (1997) 7674--7691.}
\maketitle
\begin{abstract}%
We perform a Hamiltonian reduction of spherically symmetric Einstein
gravity with a thin dust shell of positive rest mass. Three spatial
topologies are considered: Euclidean~($\BbbR^3$), Kruskal
($S^2\times\BbbR$), and the spatial topology of a diametrically identified
Kruskal ($\RPthree\setminus${}$\{$a point at infinity$\}$). For the Kruskal
and $\RPthree$ topologies the reduced phase space is four-dimensional, with
one canonical pair associated with the shell and the other with the
geometry; the latter pair disappears if one prescribes the value of the
Schwarzschild mass at an asymptopia or at a throat. For the Euclidean
topology the reduced phase space is necessarily two-dimensional, with only
the canonical pair associated with the shell surviving. A
time-reparametrization on a two-dimensional phase space is introduced and
used to bring the shell Hamiltonians to a simpler (and known) form
associated with the proper time of the shell.  An alternative
reparametrization yields a square-root Hamiltonian that generalizes the
Hamiltonian of a test shell in Minkowski space with respect to Minkowski
time.
Quantization is briefly discussed.  The discrete mass spectrum
that characterizes natural minisuperspace quantizations of vacuum
wormholes and  $\RPthree$-geons appears to persist as the geometrical
part of the mass spectrum when the additional matter degree of
freedom is added.
\end{abstract}
\pacs{Pacs:
04.20.Fy,
04.40.Nr,
04.60.Kz,
04.70.Dy
}

\narrowtext

\section{Introduction}
\label{sec:intro}

In classical general relativity, every three-manifold occurs as the
spatial topology of a globally hyperbolic vacuum spacetime. In a
canonical approach to quantum gravity, the spatial topology is frozen,
and one can ask for ground states corresponding to each
topology.\footnote{Even in a theory that permits topology change,
topologies threaded by electric or magnetic flux in source-free
Einstein-Maxwell theory (or in higher-dimensional gravity with
Kaluza-Klein asymptotic behavior) cannot evolve to Euclidean space,
if they have a net asymptotic charge. If there is a nonsingular
quantum theory of such a system, it must allow a ground state with
nonzero asymptotic charge and non-Euclidean topology. Topological geons
with half-integral angular momentum in a quantum theory of gravity
would similarly be unable to settle down to Euclidean topology.}

Spherically symmetric minisuperspaces provide simple models for the
quantization of geometries with non-Euclidean topology. The spatial
topologies consistent with spherical symmetry and asymptotic flatness are
$\BbbR^3$, the wormhole $S^2 \times \BbbR$ of the Kruskal
geometry with two asymptopias, and the $\RPthree$ geon, a manifold with a
single asymptopia obtained by removing a point from the compact manifold
$\RPthree$. This last manifold is the space acquired from Kruskal geometry
by identifying diametrically opposite points on an $U+V=\hbox{constant}$
slice, with $U$ and $V$ the usual Kruskal null coordinates \cite{topocen}.

A reduced phase space formalism for spherically symmetric vacuum
Einstein gravity in four spacetime dimensions has been considered by
several authors
\cite{bcmn,unruh,thiemann1,thiemann2,%
kuchar1,LW2,cava1,cava2,lau1,marolf-boundary,lm}.\footnote{For
extensions to related theories, including spherically symmetric
Einstein-Maxwell theory and lower-dimensional dilatonic theories, see
Refs.\
\cite{lau1,thiemann3,thiemann4,kuns1,kuns2,kuns3,%
varadarajan,BLPP,lou-win,lou-si-win}.
For discussions within the Euclidean context, see for example Refs.\
\cite{laf-stwo,WYprl,BBWY,JJHlouko3,haylo,LW1,mel-whit,%
cartei1,cartei2,oliveira}
and the references therein.}
In the present paper we add to spherically symmetric Einstein gravity an
idealized, infinitesimally thin dust shell of positive rest mass. The
equations of motion for such a shell follow easily from Israel's junction
condition formalism \cite{israel-shell,MTW,GT,barr-is}, and a number of
workers have proposed actions from which these equations can be
derived \cite{frolov,berezin88,hajicek,HKK,nakamura,dolgov,bicak-haji}. Our
main purpose is to find an action for this system by an explicit
Hamiltonian reduction, treating both the geometry and the shell as
dynamical, and retaining the full dynamics allowed by the choice of the
spatial topology.

Two issues require particular care. First, as general relativity is a
nonlinear theory, introducing a distributional source faces well
known subtleties \cite{GT}. The special case of a source concentrated on a
hypersurface of codimension~1 is fortunate, as Einstein's equations
can then be given an unambiguous distributional interpretation, and this
interpretation reduces to Israel's junction conditions when the source is a
pure delta-function on the surface \cite{GT}. However, we wish to go
further and write an {\em action principle\/} from which the field equations
would arise as variational equations. In such an action principle one
presumably needs to be able to vary the action with respect to both the
metric and the shell variables, with the variations remaining
independent in some suitable sense. This brings in not only the regularity
properties of the spacetime at the shell, but also the regularity properties
of the coordinates in which the action is written.

We will not find an action principle whose variational equations would be
fully distributionally consistent at the shell. However, the ambiguity in
our variational equations will be localized into the single equation that
results from varying the action with respect to the shell position
coordinate. When the ambiguous contribution to this equation is interpreted
as the average of its values on the two sides of the shell, as is
necessitated by consistency with the rest of the equations, we correctly
reproduce the content of Israel's junction condition formalism. At a
somewhat formal level, our action will be manifestly invariant under the
Hamiltonian version of spacetime coordinate transformations preserving
spherical symmetry.

Second, one needs to choose the falloff and boundary conditions at the
asymptopias. We shall set the asymptotic momenta to zero, but the values of
the Schwarzschild masses at the asymptopias will be left free to
emerge from the dynamics. Our spacelike hypersurfaces will not be asymptotic
to hypersurfaces of constant Minkowski time, but the foliation is
nevertheless asymptotically Minkowski in the relevant sense. In particular,
the generator of unit time translations at the infinity is the
Schwarzschild mass.

We shall find that the reduced phase space is four-dimensional with the
Kruskal and $\RPthree$ topologies, and two-dimensional with the
$\BbbR^3$ topology. With each topology, one canonical pair is
associated with the shell motion, but with the Kruskal and $\RPthree$
topologies there is also a second canonical pair, associated with the
dynamics of the geometry. In the limit where the shell is removed, this
reproduces results previously obtained in the Hamiltonian vacuum theories
with the Kruskal and $\RPthree$ topologies \cite{kuchar1,LW2}.

For the
non-Euclidean
topologies, the canonical pair associated with
the geometry disappears if one prescribes by hand the mass at one infinity
in the Kruskal topology, and the mass at the wormhole throat
in the $\RPthree$ topology. All three reduced phase spaces then become
two-dimensional, and they can be treated on an essentially equal footing.

We next introduce a formalism for reparametrizing time in a Hamiltonian
theory with a two-dimensional phase space. Applying this formalism to our
two-dimensional phase spaces, we redefine the coordinate time
to coincide with the proper time of the shell and thereby obtain a
Hamiltonian that can be given in terms of elementary functions. This
Hamiltonian is known \cite{berezin88}, but the fact that it emerges
from a minisuperspace framework is new. An alternative choice for the
coordinate time yields a Hamiltonian that generalizes to our
self-gravitating shell the familiar $\sqrt{p^2 + m^2}$ Hamiltonian of a
spherical test shell in Minkowski space.

The paper concludes with a discussion of the prospects for quantization.
Quantization of the vacuum case is revisited to emphasize choices that
lead to discrete or continuous mass spectra. The additional degree of
freedom provided by the shell does not appear to qualitatively alter these
choices.
Like the Jain-Schechter-Sorkin quantum-stabilized
Skyrmion \cite{jss},
the minisuperspace geons provide an example of field configurations
that have quantum, but not classical ground states; both are field theory
analogues of the quantum-stabilization of the hydrogen atom.
Whether the ground state of geons is an artifact of the reduction of
the degrees of freedom is, of course, an open question, but the
geometrical ground state appears to persist when the shell's
degree of freedom is added.

With one asymptotic or interior mass fixed, the implicit Hamiltonian
we obtain prior to time-reparametrization was found by Kraus and Wilczek
\cite{kraus-wilczek1,kraus-wilczek2} in the limit of a massless shell, and
it could easily be found from what they present also for the massive case.
A related reduction technique was used earlier by Fischler {\it et al\/}.\
\cite{fischler} in a minisuperspace treatment of a bubble wall, and
recently generalized by Kolitch and Eardley \cite{kol-eard-ham}.
For a flat geometry interior to the shell, our proper-time Hamiltonian has
been considered classically in Ref.\ \cite{frolov} and quantum mechanically
in Ref.\ \cite{hajicek}. For a flat geometry interior to the shell,
quantization using the square root Hamiltonian has been considered in
Refs.\ \cite{HKK,dolgov}.

Latin tensor indices $a,b,\dots$ indicate abstract spacetime
indices. We work in Planck units, $\hbar = c = G = 1$.

\section{Hamiltonian formulation for spherically symmetric geometry with a
dust shell}
\label{sec:action}

In this section we set up a Hamiltonian formulation for spherically
symmetric gravity coupled to a thin dust shell. We pay special attention to
the smoothness of the gravitational variables and to the global boundary
conditions.

\subsection{Bulk action}
\label{subsec:bulk}

A spherically symmetric spacetime metric can be locally written in the
Arnowitt-Deser-Misner (ADM) form
\begin{equation}
ds^2=-N^2dt^2 + \Lambda^2(dr+N^r dt)^2+R^2 d\Omega^2
\ \ ,
\label{metric}
\end{equation}
where $d\Omega^2$ is the metric on the unit two-sphere, and $N$, $N^r$,
$\Lambda$, and $R$ are functions of $t$ and~$r$. Issues of smoothness
and global structure will be addressed below. We denote the derivative with
respect to $t$ by overdot, and the derivative with respect to $r$ by prime.

The matter consists of a thin shell of dust, with a fixed positive rest
mass~$m$. We write the trajectory of the shell as $r={\zr }(t)$.  Denoting
by $\hat N(t)$, $\hat N^r(t)$, $\hat\Lambda (t)$, and $\hat R(t)$ the values
of $N$, $N^r$, $\Lambda$, and $R$ at $r={\zr}$,
\begin{equation}
\hat R(t) := R \biglb(t,{\zr}(t)\bigrb) \ \ \hbox{etc,}
\label{hatR}
\end{equation}
the Hamiltonian action for the shell is
\cite{kraus-wilczek1,fischler}
\begin{equation}
S_{\text{shell}} =\int dt \left( \zp\dot{\zr } - \hat N
{\sqrt{\zp^2 \hat \Lambda^{-2}+m^2}} + \hat N^r \zp\right)
\ \ ,
\label{shell-action}
\end{equation}
with $\zp$ being the momentum conjugate to~$\zr$. One can think of the shell
as a spherically symmetric cloud
of massive relativistic point particles.

The Lagrangian gravitational action for the geometry (\ref{metric}) is
obtained by integrating the Lagrangian density ${(16\pi)}^{-1} ({}^3 \! R-
K^{ab}K_{ab}+K^2) {\sqrt{-g}}$ over the
two-sphere \cite{bcmn,unruh,kuchar1,kraus-wilczek1,fischler,lund}.  After
$\dot\Lambda$ and $\dot R$ are replaced by their conjugate momenta,
\begin{mathletters}
\label{gravmomenta}
\begin{eqnarray}
\pi_\Lambda &=& - \frac{R}{N} (\dot R-N^rR')
\ \ ,
\label{gravmomenta-lambda}
\\
\pi_R &= &-\frac{\Lambda}{N} (\dot R-N^r R')
-\frac{R}{N}
\left[ \dot\Lambda -
(N^r\Lambda)' \right]
\ \ ,
\end{eqnarray}
\end{mathletters}%
the Hamiltonian bulk action for the coupled system reads
\begin{equation}
S_\Sigma =
\int dt \left[
\zp
{\dot{\zr }} + \int dr \left( \pi_\Lambda \dot\Lambda +
\pi_R \dot R - N{\zH } - N^r {\zH }_r \right)
\right]
\ \ ,
\label{bulk-action}
\end{equation}
where the super-Hamiltonian $\zH$ and the supermomentum ${\zH}_r$ are given
by
\begin{mathletters}
\begin{eqnarray}
&&{\zH } =
{\Lambda \pi^2_\Lambda \over 2 R^2}
- { \pi_\Lambda \pi_R \over R}
+ { R R'' \over \Lambda}
- { R R' \Lambda' \over \Lambda^2}
+ { R'^2 \over 2\Lambda}
- {\Lambda \over 2}
\nonumber
\\
\noalign{\smallskip}
&& \qquad \; + {\sqrt{\zp^2\hat\Lambda^{-2}+m^2}}
\,
\delta (r-{\zr } )
\ \ ,
\label{superham}
\\
\noalign{\smallskip}
&&{\zH }_r = \pi_R R' - \pi'_\Lambda\Lambda - \zp\delta (r-\zr )
\ \ .
\label{supermom}
\end{eqnarray}
\end{mathletters}%
We shall first discuss the smoothness of the gravitational variables, and
then the boundary terms to be added to the bulk action.

\subsection{Smoothness}
\label{subsec:smoothness}

In the presence of a smooth matter distribution, one can assume the
spacetime metric to be
smooth~($C^\infty$). In the idealized case of an
infinitesimally thin shell, the metric can be chosen to be continuous
but not, in general, differentiable across the shell
\cite{israel-shell,MTW,GT,barr-is}. In the particular case of a
spherically symmetric dust shell, Einstein's equations imply that the
extrinsic curvature of the shell history is discontinuous
both in its angular components and in its component along the shell
four-velocity. If the metric is taken continuous, we must therefore
accommodate discontinuities in $R'$ and in at least
some\footnote{\label{foot:Delta}By continuity of the metric,
$\hat R (t)$ is well defined for all~$t$. Taking the total time derivative
of (\ref{hatR}) shows that $\Delta {\dot R} = - \dot{\zr} \Delta R'$, where
$\Delta$ denotes the discontinuity across the
shell. Similarly for $\Lambda$, $N$, and~$N^r$. Continuity of $\Lambda'$,
$N'$, and $(N^r)'$ would therefore imply that the extrinsic curvature of
the shell history is discontinuous only in its angular
components.} of $\Lambda'$, $N'$, and~$(N^r)'$. We would
like both the action (\ref{bulk-action}) and its local variations to be
well defined, and such that the resulting variational equations are
equivalent to Einstein's equations with a dust shell.

To proceed, we assume that the gravitational variables are smooth
functions of~$r$, with the exception that $N'$, $(N^r)'$, $\Lambda'$, $R'$,
$\pi_\Lambda$, and $\pi_R$ may have finite discontinuities at isolated
values of~$r$, and that the coordinate loci of the discontinuities may be
smooth functions of~$t$. It will be shown that the resulting variational
principle is satisfactory in the above sense, provided one of the
variational equations is interpreted as the average of a discontinuous
quantity over the two sides of the shell.\footnote{Because  the constraint
equations enforce smoothness of the metric outside the  shell, our
differentiability assumptions can probably be relaxed.} All the terms under
the $r$-integral in the action (\ref{bulk-action}) are well defined in the
distributional sense. The most singular contributions are the explicit
matter delta-contributions in the constraints, and the implicit
delta-functions in $R''$ and~$\pi'_\Lambda$. All these delta-functions are
multiplied by continuous functions of~$r$. The remaining terms are at worst
discontinuous in~$r$. The action is therefore well defined.

Local independent variations of the action with respect to the
gravitational and matter variables give the
constraint equations
\begin{mathletters}
\label{metric-constr-eoms}
\begin{eqnarray}
{\zH} &=& 0
\ \ ,
\label{metric-constr-eoms-ham}
\\
{\zH}_r &=& 0
\ \ ,
\end{eqnarray}
\end{mathletters}%
and the dynamical equations
\begin{mathletters}
\label{metric-dyn-eoms}
\begin{eqnarray}
{\dot \Lambda}
&=&
N \left( {\Lambda \pi_\Lambda \over R^2}
    - {\pi_R \over R} \right)
+ {\left( N^r \Lambda \right)}'
\ \ ,
\label{Lambda-eom}
\\
{\dot R}
&=&
- {N \pi_\Lambda \over R}
+ N^r R'
\ \ ,
\label{R-eom}
\\
{\dot \pi}_\Lambda
&=&
{N \over 2}
\left[
- {\pi_\Lambda^2 \over R^2}
- \left({R' \over \Lambda}\right)^2
+ 1
+ {2 {\zp}^2 \delta(r-{\zr})
\over
{\hat \Lambda}^3 {\sqrt{\zp^2 \hat \Lambda^{-2}+m^2}}}
\right]
- {N' R R' \over \Lambda^2}
+ N^r \pi_\Lambda'
\ \ ,
\label{PLambda-eom}
\\
{\dot \pi}_R
&=&
N
\left[
{\Lambda \pi_\Lambda^2 \over R^3}
- {\pi_\Lambda \pi_R \over R^2}
- \left({R' \over \Lambda}\right)'
\right]
- \left( {N' R \over \Lambda}\right)'
+ {\left( N^r \pi_R \right)}'
\ \ ,
\label{PR-eom}
\\
{\dot {\zr}}
&=&
{ {\hat N} {\zp}
\over
{\hat \Lambda}^2 {\sqrt{\zp^2 \hat \Lambda^{-2}+m^2}}}
- {\hat N}^r
\ \ ,
\label{rhat-eom}
\\
{\dot {\zp}}
&=&
{ {\hat N} {\hat \Lambda}' {\zp}^2
\over
{\hat \Lambda}^3 {\sqrt{\zp^2 \hat \Lambda^{-2}+m^2}}}
- {\hat N}' {\sqrt{\zp^2 \hat \Lambda^{-2}+m^2}}
+  {\zp} {\widehat {(N^r)'}}
\ \ .
\label{rhatmom-eom}
\end{eqnarray}
\end{mathletters}%
With the exception
of~(\ref{rhatmom-eom}), all the equations
(\ref{metric-constr-eoms}) and (\ref{metric-dyn-eoms}) are
well defined in a distributional sense.\footnote{The constraint equations
(\ref{metric-constr-eoms}) contain explicit delta-functions in $r$ from the
matter contribution and implicit delta-functions in
$R''$ and~$\pi_\Lambda'$. The right-hand sides
of (\ref{Lambda-eom}) and (\ref{R-eom}) contain at worst finite
discontinuities, and the right-hand sides of (\ref{PLambda-eom}) and
(\ref{PR-eom}) contain at worst delta-functions. This is consistent with the
left-hand sides of (\ref{Lambda-eom})--(\ref{PR-eom}), because the
loci of nonsmoothness in $\Lambda$, $R$, $\pi_\Lambda$ and $\pi_R$ may
evolve smoothly in~$t$. Note that both the explicit matter delta-functions
and the implicit delta-functions in $R''$ and $\pi_\Lambda'$ are multiplied
by continuous functions of~$r$.} What needs to be examined is the
consistency and dynamical content of the well-defined equations, and the
interpretation of  the single troublesome equation~(\ref{rhatmom-eom}).

As a preliminary, consider the variation of the matter action
$S_{\text{shell}}$ (\ref{shell-action}) with respect to the metric. {}From
the definition of the stress-energy tensor,
\begin{equation}
\delta_g S_{\text{shell}} = \casehalf \int \sqrt{-g}\, d^4x \, T^{ab} \,
\delta(g_{ab})
\ \ ,
\end{equation}
and the equation of motion~(\ref{rhat-eom}), we find that the surface
stress-energy tensor of the shell
[Ref.\ \cite{MTW}, equation (21.163)]
takes the form
\begin{equation}
S^{ab} = { m \over 4 \pi {\hat R}^2 } \, u^a u^b
\ \ ,
\label{dust-sem}
\end{equation}
where $u^a$ is the four-velocity of the shell, normalized in the usual way
$u^a u_a=-1$. This confirms that the shell indeed consists of pressureless
dust, with surface energy density $m/(4 \pi {\hat R}^2)$ and total rest
mass~$m$.

Also, recall that the full content of the Einstein equations at the shell
is encoded in Israel's junction conditions \cite{israel-shell,MTW}. We shall
refer to the two sides of the shell as the ``right-hand side" and the
``left-hand side", in view of the Penrose diagram in which two partial
Kruskal diagrams are joined to each other along the shell
trajectory.  Israel's junction conditions then read
\begin{equation}
- 8\pi ( S_{ab} - \casehalf h_{ab} S ) = K_{ab}^+ - K_{ab}^-
\ \ ,
\label{junction-cond}
\end{equation}
where $n_a$ is the right-pointing unit normal to the shell history,
$h_{ab} = g_{ab} - n_a n_b$ is the projector to this history,
$K_{ab} = h^c{}_a h^d{}_b \nabla_c n_d$ is the extrinsic curvature tensor,
and the signs $\pm$ refer respectively to the right and left sides of the
shell. With~(\ref{dust-sem}), and with Kruskal geometries of masses
$M_\pm$ on the two sides of the shell, the angular components
of (\ref{junction-cond}) read
\begin{equation}
- {m \over {\hat R} }
=
\epsilon_+ \sqrt{ \left(d{\hat R} \over d\tau\right)^2
+ 1 - {2 M_+\over{\hat R}}}
\ - \
\epsilon_- \sqrt{ \left(d{\hat R} \over d\tau\right)^2
+ 1 - {2 M_-\over{\hat R}}}
\ \ ,
\label{shell-full-eom}
\end{equation}
where $\tau$ is the shell's proper time. $\epsilon_+=1$ ($\epsilon_-=1$) if,
when viewed from the geometry right (left) of the shell, the shell is in the
right-hand-side exterior region of the Kruskal diagram, or if the shell is
in the white-hole region and moving to the right, or if the shell is in the
black-hole region and moving to the left. Otherwise $\epsilon_+=-1$
($\epsilon_-=-1$). It can be verified that the shell motion is completely
determined by the single equation (\ref{shell-full-eom}) and the vacuum
Einstein equations away from the shell. In particular, these equations imply
that the tangential component of (\ref{junction-cond}) is satisfied.
A more explicit discussion can be found in Ref.\ \cite{bicak-haji}.

Now, away from the shell,
equations (\ref{metric-constr-eoms})
and (\ref{metric-dyn-eoms}) are well known to be equivalent to Einstein's
equations. At the shell, the constraints
(\ref{metric-constr-eoms}) read
\begin{mathletters}
\label{delta-constraints}
\begin{eqnarray}
&&\Delta R'  = - { \sqrt{{\zp}^2 + m^2 {\hat \Lambda}^2}
\over
{\hat R}}
\ \ ,
\label{delta-ham}
\\
&&\Delta \pi_\Lambda = - {\zp \over {\hat\Lambda}}
\ \ ,
\label{delta-mom}
\end{eqnarray}
\end{mathletters}
where we have adopted the notation (cf.\ footnote \ref{foot:Delta})
\begin{equation}
\Delta f := \lim_{\epsilon\to 0_+}
\left[ f(\zr + \epsilon)
- f(\zr - \epsilon) \right]
\ \ ,
\end{equation}
identifying $r>\zr$ ($r<\zr$) as the right (left) side of the shell.
Using (\ref{rhat-eom}) and~(\ref{dust-sem}), one finds that (\ref{delta-ham})
is equivalent to~(\ref{shell-full-eom}). Our
Hamiltonian equations away from the shell and the Hamiltonian constraint
(\ref{metric-constr-eoms-ham}) at the shell therefore form a system that is
equivalent to the correct dynamics for the shell. When these equations hold,
it can be verified that equation (\ref{delta-mom}) is proportional to
(\ref{delta-ham}) by the nonsingular factor ${\hat R} ( {\dot{\zr}} + {\hat
N}^r ) / {\hat N}$, and the momentum constraint thus contains no new
information. Similarly, it can be verified that the delta-parts in
(\ref{PLambda-eom}) and (\ref{PR-eom}) reduce to identities and contain no
new information. Finally, (\ref{Lambda-eom}) and (\ref{R-eom}) contain no
delta-functions, and thus no new information, at the shell.

The single remaining equation of motion is~(\ref{rhatmom-eom}). If the
geometry were smooth at the shell, equations (\ref{rhat-eom}) and
(\ref{rhatmom-eom}) would by construction be equivalent to the geodesic
equation for the shell, as can indeed be explicitly verified. If the
ambiguous spatial derivative terms in (\ref{rhatmom-eom}) are evaluated on
the left (right) side of the shell, (\ref{rhatmom-eom}) thus implies the
geodesic equation for the shell in the geometry on the left (right).
However, these two geodesic equations are mutually inconsistent, and the
shell motion implied by the rest of the equations is not geodesic in either
of the two geometries. Instead, the rest of the equations imply that the
left-hand side of (\ref{rhatmom-eom}) is equal to the {\em average\/} of the
right-hand side over the two sides of the shell. (For the generalization of
this observation to nonspherical dust shells, see Exercise 21.26 in Ref.\
\cite{MTW}.) Therefore, if the ill-defined right-hand side of
(\ref{rhatmom-eom}) is given this average interpretation, our Hamiltonian
formalism reproduces Einstein's equations for the dust shell.

We are not aware of an a priori justification of the averaged interpretation
of the right-hand side of~(\ref{rhatmom-eom}). This interpretation is merely
forced on us by the rest of the variational equations. In a strict sense, we
therefore regard the variational principle as inconsistent, and the averaged
interpretation of (\ref{rhatmom-eom}) as put in by hand. Nevertheless, we
shall proceed with this variational principle. It will be seen
in section \ref{sec:reduction} that the Hamiltonian reduction can be carried
through with no apparent inconsistency.

One check on the consistency of the formalism is that the Poisson
brackets of our constraints can be verified to obey the radial hypersurface
deformation algebra \cite{teitel-dirac}, as in the absence of the
shell. If we denote by
$\zN (r)$ and ${\zN }^r(r)$ smooth smearing functions of
compact support, the algebra has the form
\begin{mathletters}
\label{pb-algebra}
\begin{eqnarray}
\left\{ \int dr\, {\zN }_1 \zH , \hspace{1mm} \int dr \,
{\zN }_2\zH\right\}
&=& \int dr\, \left( {\zN }_1 {\zN }'_2 - {\zN }_2 {\zN }'_1\right)
\Lambda^{-2}\zH_r
\ \ ,
\\
\left\{ \int dr\, {\zN }^r \zH_r , \hspace{1mm} \int dr\, {\zN }\zH
\right\} &=& \int dr\, {\zN }^r {\zN }'\zH
\ \ ,
\\
\left\{ \int dr\, {\zN }_1^r \zH_r , \hspace{1mm} \int dr\, {\zN }_2^r\zH_r
\right\}
&=&
\int dr\, \left[ {\zN }_1^r ({\zN }_2^r)' -
{\zN }_2^r({\zN }_1^r)'\right]\zH_r
\ \ .
\end{eqnarray}
\end{mathletters}

\subsection{Asymptopias and boundary terms}
\label{subsec:asymptopias}

We now turn to the global properties of the geometry. In this section we
take the spatial topology to be that of the extended Schwarzschild
geometry,  $S^2 \times \BbbR = S^3 \setminus${}$\{$two points$\}$, the
omitted points being associated with asymptotically flat
asymptopias. The spatial topologies $\RPthree\setminus${}$\{$a point at
infinity$\}$ and $\BbbR^3$ will be discussed respectively
in sections \ref{sec:geon} and~\ref{sec:flat-interior}.

At a general level, restricting the asymptotic behavior of
an asymptotically flat system allows one to fix the momentum, angular
momentum, and mass at spatial infinity. In a quantum theoretic
context, to restrict in this way the asymptotic behavior of the operator
$^3\hat g_{ab}$ and its conjugate momentum $\hat\pi^{ab}$ is equivalent to
restricting the state space to an eigensubspace of fixed total
momentum, angular momentum, or mass. In our particular case of spherical
symmetry, the angular momentum is necessarily zero. It would be
consistent with spherical symmetry to allow a nonzero momentum at infinity
(in the classical framework, this would mean allowing boosted Schwarzschild
solutions), but for our purposes this freedom does not appear significant,
and we shall set the momentum at infinity to zero.
We shall, however, retain the freedom associated with the system's total
mass.

We take the coordinate $r$ to have the range $-\infty < r < \infty$. At
the asymptopias $r\to\pm\infty$, we introduce the falloff
\begin{mathletters}
\label{l-r}
\begin{eqnarray}
\Lambda(t,r)
&=&
1
+ O^\infty \! \left({|r|}^{-{3 \over 2} -\fallpar}\right)
\ \ ,
\label{l-r-Lambda}
\\
R(t,r)
&=&
{|r|} + O^\infty \! \left({|r|}^{-{1 \over 2} -\fallpar}\right)
\ \ ,
\label{l-r-R}
\\
\pi_{\Lambda}(t,r)
&=&
\sqrt{2M_\pm {|r|}}
+ O^\infty \! \left({|r|}^{-\fallpar}\right)
\ \ ,
\label{l-r-PLambda}
\\
\pi_{R}(t,r)
&=&
\sqrt{ {M_\pm \over 2 {|r|} }}
+ O^\infty \! \left({|r|}^{-1 -\fallpar}\right)
\ \ ,
\label{l-r-PR}
\\
N(t,r) &=&
1
+ O^\infty \! \left({|r|}^{-\fallpar}\right)
\ \ ,
\label{l-r-N}
\\
N^r(t,r) &=&
\pm \sqrt{ {2M_\pm \over {|r|} }}
+ O^\infty \! \left({|r|}^{-{1 \over 2} -\fallpar}\right)
\ \ ,
\label{l-r-Nr}
\end{eqnarray}
\end{mathletters}%
where $M_\pm(t)$ are positive-valued functions of~$t$, and $\fallpar$ is
a positive parameter that can be chosen at will. $O^\infty$~indicates a
quantity that is bounded at infinity by a constant times its argument,
with the corresponding behavior for its derivatives.

It is straightforward to verify that the falloff (\ref{l-r}) is consistent
with the constraints and preserved in time by the dynamical equations. When
the equations of motion hold, $M_\pm$ are independent of~$t$, and their
values are just the Schwarzschild masses at the two asymptopias.
Using~(\ref{shell-full-eom}), it is easy to show that the existence of two
asymptotically flat infinities implies that both asymptotic Schwarzschild
masses in the classical solutions are necessarily positive. The
assumption $M_\pm(t)>0$ in (\ref{l-r}) does therefore not exclude any
solutions.

The falloff (\ref{l-r}) is not consistent with the conventional falloffs
(see, for example, Refs.\ \cite{kuchar1,beig-om}) in which the
hypersurfaces of constant $t$ are asymptotic to hypersurfaces of constant
Killing time when the equations of motion hold. Instead, the falloff
(\ref{l-r}) is asymptotic to the ingoing spatially flat coordinates
\cite{painleve,gullstrand,israel-flat}, individually near each asymptopia.
When $M_\pm$ are constants and all the $O^\infty$-terms
vanish, (\ref{l-r}) yields the Schwarzschild metric in the ingoing spatially
flat coordinates, separately for $r>0$ and $r<0$. Our reason for adopting
(\ref{l-r}) is that the spatially flat gauge will prove useful in the
Hamiltonian reduction in section \ref{sec:reduction} \cite{kraus-wilczek1}.

In a variational principle that does not fix the values of~$M_\pm$, the bulk
action (\ref{bulk-action}) must be amended by a boundary action. With our
falloff~(\ref{l-r}), the spatial metric approaches flatness at
$r\to\pm\infty$ so fast that the variations of $R$ and $\Lambda$ give rise
to no boundary terms from the infinities. The only nontrivial boundary
term arises from integrating by parts the term $\int dt \int
dr\, N^r\Lambda (\delta\pi_\Lambda )'$, associated with the momentum
constraint.  This boundary term is
canceled if we add to the bulk action
(\ref{bulk-action}) the boundary action
\begin{equation}
S_{\partial\Sigma} = - \int dt \, \left( M_+ + M_- \right)
\ \ .
\label{S-psigma1}
\end{equation}
The generator of unit time translations at the infinities is therefore still
the Schwarzschild mass, despite the unconventional falloff.

\section{Reduced phase space formulation}
\label{sec:reduction}

In the absence of the shell, the Hamiltonian reduction of our theory with
a technically different but qualitatively similar falloff at the two
asymptopias was discussed in Ref.\ \cite{kuchar1}. When the asymptotic
masses are not fixed, it was found that the reduced phase space is
two-dimensional, whereas if one asymptotic mass is fixed, the reduced phase
space has dimension zero. As the shell brings in one new canonical pair but
no new constraints, one expects that the reduced phase space of our theory
is four-dimensional when the asymptotic masses are not fixed, and
two-dimensional if one asymptotic mass is fixed. In this section we shall
verify this expectation by an explicit Hamiltonian reduction.

\subsection{Gauge transformations and the Hamiltonian reduction formalism}

In the Hamiltonian theory formulated in section~\ref{sec:action}, the
variables $\left(\Lambda, R, \pi_\Lambda, \pi_R, \zr, \zp \right)$
constitute a canonical chart on the phase space~${\cal S}$, while $N$ and
$N^r$ act as Lagrange multipliers enforcing the constraints. As the Poisson
bracket algebra (\ref{pb-algebra}) of the constraints closes, we have a
first class constrained system \cite{henn-teit-book}.

Let $\Gamma$ denote the constraint hypersurface (\ref{metric-constr-eoms})
in~${\cal S}$. We take gauge transformations to mean the transformations on
$\Gamma$ generated by the constraints.\footnote{See, for example, Ref.\
\cite{ash-geroch-rev}. Note that this is distinct from, although closely
related to, the gauge transformations that act on the histories on which the
action is defined
\cite{henn-teit-book,teitel-qmgrav1,teitel-qmgrav2}.} Denoting the smearing
functions by $\zN(r)$ and ${\zN }^r(r)$ as in~(\ref{pb-algebra}), the
smeared Hamiltonian constraint transforms an initial data set
$\left(\Lambda, R, \pi_\Lambda, \pi_R, \zr, \zp \right)\in\Gamma$
by the time evolution associated with~$\zN$, and the smeared momentum
constraint transforms the initial data set by the spatial diffeomorphism
associated with~${\zN}^r$. The smearing functions must fall off so fast
that the transformations become trivial at the infinities and the  falloff
(\ref{l-r}) is preserved.\footnote{One could consider an extended phase
space that contains $N$ and $N^r$ as new coordinates and their conjugates
$\pi_N$ and $\pi_{N^r}$ as new momenta. We shall, however, not need this
extension.}

By definition, the reduced phase space ${\bar\Gamma}$ consists of the
equivalence classes in $\Gamma$ under gauge transformations.
The symplectic form $\omega$ on~${\cal S}$,
\begin{equation}
\omega :=
\extder\zp \wedge \extder\zr
+ \int dr \left(\extder \pi_\Lambda \wedge \extder \Lambda
+ \extder \pi_R \wedge \extder R\right)
\ \ ,
\label{omega}
\end{equation}
induces a symplectic form ${\hat\omega}$ on~${\bar\Gamma}$. Here,
and from now on, $\extder$ denotes the exterior derivative on the
(functional) spaces in question.

We wish to implement this Hamiltonian reduction, finding ${\hat\omega}$ in
an explicit symplectic chart on~${\bar\Gamma}$. Our implementation will
consist of the following three steps:

1. Consider first~$\Gamma$. At the shell, we have
already seen that the full content of the constraints is encoded in
equations~(\ref{delta-constraints}). Away from the shell, the constraints
can be solved explicitly for the gravitational momenta as
\cite{kraus-wilczek1,fischler}
\begin{mathletters}
\label{pi-sol}
\begin{eqnarray}
\pi_\Lambda
&=&
R {\sqrt{(R'/\Lambda)^2-1+2M_\pm/R}}
\ \ ,
\label{pi-lambda-sol}
\\
\pi_R
&=&
{\Lambda
\left[
(R/\Lambda) (R'/\Lambda)'
+ (R'/\Lambda)^2
- 1
+M_\pm/R
\right]
\over
\sqrt{(R'/\Lambda)^2-1+2M_\pm/R}
}
\ \ ,
\label{pi-R-sol}
\end{eqnarray}
\end{mathletters}%
with the upper (lower) signs holding respectively for $r>\zr$ ($r<\zr$).
We have chosen the sign of the square root in (\ref{pi-sol}) so as to agree
with the falloff~(\ref{l-r}). This choice will lead to a reduction that will
cover the black hole interior but not the white hole interior.

2. To pass from $\Gamma$ to~${\bar\Gamma}$, we choose a gauge: we
specify in $\Gamma$ a hypersurface $\transsurface$ that is transversal to
the gauge orbits, so that each point in (an open subset of) ${\bar\Gamma}$
has a unique representative in~$\transsurface$. This defines an isomorphism
between $\transsurface$ and (the open subset of)~${\bar\Gamma}$. In
order to choose the gauge in practice, we note that away from the
shell, a point $\left(\Lambda, R,
\pi_\Lambda, \pi_R, \zr, \zp\right)\in\Gamma$
is an initial data set for
the {\em vacuum\/} Einstein equations with spherical symmetry. Any vacuum
initial data set has a unique time evolution, and, by Birkhoff's theorem,
the resulting subspacetimes left and right of the shell are isometric to
regions of two Kruskal spacetimes with the respective masses $M_-$
and~$M_+$. A solution to the constraint equations can thus be regarded as
two parametrized partial spacelike hypersurfaces in the two Kruskal
spacetimes, joining appropriately at the shell. In this picture, a gauge
choice means making a particular choice for these two partial spacelike
hypersurfaces in the two Kruskal spacetimes, in a way that joins
appropriately at the shell and is compatible with the falloff at the
infinities.

3. To find the symplectic form ${\hat\omega}$ on~${\bar\Gamma}$, it is
convenient first to find the corresponding Liouville form. Recall that
on~${\cal S}$, the Liouville form corresponding to our canonical chart
$\left(\Lambda, R, \pi_\Lambda, \pi_R, \zr, \zp \right)$ is
\begin{equation}
\theta :=
\zp \extder\zr
+ \int dr \left(\pi_\Lambda \extder \Lambda + \pi_R \extder R\right)
\ \ .
\label{theta}
\end{equation}
Pulling $\theta$ back to $\transsurface$ yields on $\transsurface$ the
Liouville form~${\hat\theta}_\transsurface$, and
${\hat\omega}_\transsurface:=\extder {\hat\theta}_\transsurface$ is the
symplectic form on on $\transsurface$ that corresponds to~${\hat\omega}$
on (the isomorphic open subset of)~${\bar\Gamma}$.

In view of the description of the gauge choice in step~2, a technical point
in step 3 arises from the fact that although $M_\pm$ are constants in the
time evolution of a given initial data set, they are not constants as
functions on~$\transsurface$, and their exterior derivatives
may contribute to the pullback of $\theta$~(\ref{theta}). Put differently, a
generic path in ${\bar\Gamma}$ need not correspond to a partial foliation of
a single Kruskal geometry on either side of the shell.

To complete steps 2 and~3, we need to specify the gauge. This will
be described next.

\subsection{Gauge choice}
\label{subsec:gauge-choice}

Our gauge choice involves taking the intrinsic metric on the
spacelike hypersurface to be flat, with the exception of certain
transition regions that are eventually taken to be vanishingly narrow.
The possible locations for the transition regions depend on whether the
shell trajectory is visible to the right-hand-side future null infinity, the
left-hand-side future null infinity, or neither.\footnote{This last case
occurs when
$\epsilon_+=-1$ and  $\epsilon_-=1$ in~(\ref{shell-full-eom}). The spacetime
has two bifurcation spheres, and the shell passes between them, remaining at
all times behind the white-hole and black-hole horizons of each
infinity \cite{CaldChaGi}.
We are grateful to Eric Poisson for discussions on this case.}
We now make the simplifying assumption that part of the shell trajectory is
visible to one future null infinity, and we take this infinity to be on the
right. This is arguably the situation of physical interest
for an observer in the asymptotically flat region.

Thus, fix an initial data set (a point in~$\Gamma$), and consider the
classical spacetime that is its time evolution. We assume that in this
spacetime, the  shell trajectory intersects the right-hand-side exterior
region in the Kruskal geometry right of the shell. The
shell equation of motion (\ref{shell-full-eom}) then implies $M_+>M_-$, and
the trajectory intersects the right-hand-side exterior region also in the
Kruskal geometry left of the shell. It follows that $\epsilon_-=1$ on all of
the trajectory, whereas $\epsilon_+=1$ when ${\hat R}$ is sufficiently large
(in particular, when ${\hat R}> 2M_+$) but
$\epsilon_+=-1$ as ${\hat R}\to0$.

On this spacetime, we introduce two local charts, $C_1$ and $C_2$, as
follows:

Suppressing the angles, let the coordinates in the chart $C_1$ be
$(t_1,r_1)$, with $r_1>0$. The metric reads
\begin{mathletters}
\label{sflat1-chart}
\begin{eqnarray}
ds^2 &=&
-dt_1^2 + \left( dr_1 + {\sqrt{\frac{2M_-}{r_1}}} \, dt_1\right)^2 +
r_1^2d\Omega^2 ,
\quad  0 < r_1 \leq \zr - \shellpar
\ \ ,
\label{sflat1-chart-left}
\\
ds^2 &=&
-dt_1^2 + \left( dr_1 + {\sqrt{\frac{2M_+}{r_1}}} \, dt_1\right)^2
+r_1^2d\Omega^2,
\quad   \zr \le r_1
\ \ ,
\label{sflat1-chart-right}
\end{eqnarray}
\end{mathletters}%
where $\shellpar$ is a positive parameter. The two metrics shown in
(\ref{sflat1-chart}) are the ingoing right-hand-side spatially flat charts
in Kruskal manifolds with the respective masses $M_-$ and~$M_+$
\cite{painleve,gullstrand,israel-flat}. If taken individually for
$0<r_1<\infty$ and $-\infty < t_1 < \infty$, each of these two metrics would
cover the upper right half (that is, the right-hand-side exterior and
the black hole interior) in the respective full Kruskal diagrams. With the
domains indicated in~(\ref{sflat1-chart}), the combined chart is spatially
flat with mass $M_-$ for $r_1\le \zr-\shellpar$, and spatially flat with
mass $M_+$ for $r_1 \ge \zr$. The chart in the transition region
$\zr-\shellpar \le r_1 \le \zr$ will be specified below.

Let the coordinates in the chart $C_2$ be $(t_2,r_2)$, with
$r_2<0$. The metric reads
\begin{equation}
ds^2 =
-dt_2^2 + \left( - dr_2 + {\sqrt{\frac{2M_-}{|r_2|}}} \, dt_2\right)^2
+r_2^2d\Omega^2,
\quad   r_2 < 0
\ \ .
\label{sflat2-chart}
\end{equation}
We identify $C_2$ as the ingoing left-hand-side spatially flat chart in
a Kruskal manifold with mass~$M_-$, with $r_2\to-\infty$ giving the infinity
on the left.  If $-\infty < t_2 < \infty$, the metric (\ref{sflat2-chart})
covers the upper left half (that is, the left-hand-side exterior and the
black hole interior) in the Penrose diagram of this Kruskal manifold. On our
spacetime, $C_2$ covers the corresponding regions left of the shell.

Now, consider our initial data set as a parametrized spacelike hypersurface
$\Sigma_0$ in this spacetime. By the falloff~(\ref{l-r}), $\Sigma_0$ is
asymptotic at $r\to\infty$ to a constant $t_1$ hypersurface $\Sigma_1$ in
the chart~$C_1$, with $r$ being asymptotic to~$r_1$. Similarly, $\Sigma_0$
is asymptotic at $r\to-\infty$ to a constant $t_2$ hypersurface $\Sigma_2$
in the chart~$C_2$, with $r$ being asymptotic to~$r_2$. Without loss of
generality, we can take $\Sigma_1$ and $\Sigma_2$ to be respectively the
hypersurfaces $t_1=0$ and $t_2=0$. We assume that $\Sigma_1$ and $\Sigma_2$
intersect, and that they do so left of the shell, in the black hole
interior in the left-hand-side Kruskal geometry.\footnote{This assumption
is a further restriction on the initial data. Qualitatively, it tells how
``early" or ``late" the asymptotic ends of $\Sigma_0$ may be with respect
to each other and the shell trajectory.}
The value of $R$ at the intersection (where $R=r_1=-r_2$) is denoted
by~$\rho$. Note that $\rho$ can be regarded as a piece of gauge-invariant
information in our initial data set.

Let ${\hat \Sigma}_0$ be the hypersurface consisting of $\Sigma_1$ for
$r_1\ge\rho$ and $\Sigma_2$ for $r_2\le-\rho$. ${\hat \Sigma}_0$~is not
smooth, but has a corner (a sharp ridge)
at $r_1=-r_2=\rho$. We choose a positive parameter~$\ridgepar$, and we take
$\ridgepar$ and $l$ so small that
$(1+\ridgepar)\rho < \min (2M_-,\zr -\shellpar)$. We now deform ${\hat
\Sigma}_0$ in the regions $-(1+\ridgepar)\rho \le r_2 \le -\rho$ and $\rho
\le r_1 \le (1+\ridgepar)\rho$, in a way
specified below, so that the deformed hypersurface ${\tilde \Sigma}_0$
becomes a smooth, parametrized hypersurface, with a parameter $r$ that
coincides with $r_1$ for $r \ge (1+\ridgepar)\rho$ and with $r_2$ for $r \le
-(1+\ridgepar)\rho$. The canonical data on ${\tilde \Sigma}_0$ is by
construction gauge equivalent to our original initial data, and it becomes
uniquely determined after we specify ${\tilde \Sigma}_0$ in the transition
regions $|r| \le (1+\ridgepar)\rho$ and $\zr-\shellpar \le r \le \zr$.
As our gauge choice, we take canonical data on ${\tilde \Sigma}_0$ as the
representative from the gauge equivalence class of our original initial
data.

\subsection{Liouville form and the reduced Hamiltonian theory}
\label{subsec:Liouville-form}

We now find the Liouville form ${\hat\theta}_\transsurface$ by pulling the
Liouville form $\theta$~(\ref{theta}) back to the
transversal surface~$\transsurface$. This means that we need to evaluate the
right-hand side of (\ref{theta}) when the constraints and our gauge
condition hold.

Outside the transition regions $|r| \le
(1+\ridgepar)\rho$ and $\zr-\shellpar \le r \le \zr$, our gauge reads
\begin{mathletters}
\label{simpleflatgauge}
\begin{eqnarray}
R(r) &=& |r|
\ \ ,
\\
\Lambda(r) &=& 1
\ \ ,
\end{eqnarray}
\end{mathletters}%
with the gravitational momenta given by~(\ref{pi-sol}).
With~(\ref{simpleflatgauge}), $\extder R$ and $\extder\Lambda$ vanish. The
only contributions to the integral in (\ref{theta}) therefore come from the
transition regions. We evaluate these contributions in appendices
\ref{app:ridge-trans} and~\ref{app:shell-trans}, specifying the gauge
in the transition regions and finally passing to the limit where the
parameters $\shellpar$ and $\ridgepar$ vanish. {}From
(\ref{ridge-trans-cont}) and~(\ref{shell-trans-cont}), we find
\begin{equation}
{\hat\theta}_\transsurface = p_\rho \extder \rho + p \extder \zr
\ \ ,
\end{equation}
where
\begin{mathletters}
\label{red-moms}
\begin{eqnarray}
p_\rho
&:=&
\rho
\ln
\left(
\sqrt{2 M_-} + \sqrt{\rho}
\over
\sqrt{2 M_-} - \sqrt{\rho}
\right)
- 2 \sqrt{2M_-\rho}
\ \ ,
\label{rho-mom}
\\
p
&:=&
\sqrt{2M_-\zr} - \sqrt{2M_+\zr}
+ \zr \ln
\left(
\zr + \zp + \sqrt{{\zp}^2+m^2} + \sqrt{2M_+\zr}
\over
\zr + \sqrt{2M_-\zr}
\right)
\ \ ,
\label{red-shell-mom}
\end{eqnarray}
\end{mathletters}%
with $\zp$ being a solution to
\begin{equation}
M_+ - M_- = \sqrt{\zp^2+m^2} + {m^2 \over 2 \zr}
- \zp \sqrt{2M_+\over\zr}
\ \ .
\label{diffMmom}
\end{equation}
Equation (\ref{diffMmom}) has been obtained by eliminating $R'_-$ from
(\ref{gdelta-mom}) and (\ref{gRprimeminus}).

The reduction is thus complete. The functions $(\rho, p_\rho, \zr, p)$
provide a local
canonical chart on the reduced phase space~${\bar\Gamma}$, and
equations (\ref{red-moms}) and~(\ref{diffMmom}) determine $M_+$ and $M_-$ as
functions in this canonical chart.
The Hamiltonian, read off from~(\ref{S-psigma1}), is
\begin{equation}
h := M_+ + M_-
\ \ ,
\label{h-def}
\end{equation}
and the reduced action reads
\begin{equation}
S = \int dt \, ( p_\rho {\dot\rho} + p {\dot{\zr}} - h)
\ \ .
\label{red-action-rho}
\end{equation}
As anticipated, ${\bar\Gamma}$ has dimension~4.

\subsection{Dynamics in the reduced theory}
\label{subsec:red-dynamics}

For understanding the dynamical content of the reduced theory, it is
useful to introduce the new canonical chart $(M_-, P_-, \zr, p)$, defined
by (\ref{rho-mom}) and
\begin{equation}
P_-
:=
4M_-
\ln
\left(
\sqrt{2 M_-} + \sqrt{\rho}
\over
\sqrt{2 M_-} - \sqrt{\rho}
\right)
- 4 \sqrt{2M_-\rho}
\ \ .
\label{Pminus}
\end{equation}
The new action reads
\begin{equation}
S = \int dt \, ( P_- {\dot M}_- + p {\dot{\zr}} - h)
\ \ ,
\label{Mminus+shell-action}
\end{equation}
where the Hamiltonian $h(\zr,p,M_-)$ is determined by equations
(\ref{red-shell-mom})--(\ref{h-def}). In this chart, it is immediate that
both $M_-$ and $M_+$ are constants of motion. It is straightforward to
verify that the equations of motion for the shell variables are equivalent
to~(\ref{shell-full-eom}), and thus yield the correct dynamics, provided
$t$ is identified as the coordinate $t_1$ in the spatially flat chart
(\ref{sflat1-chart-right}) right of the shell. The two solutions of
(\ref{diffMmom}) for $\zp$ correspond to $\epsilon_+=\pm1$
in~(\ref{shell-full-eom}), whereas $\epsilon_-=1$ always by virtue of the
global assumptions made above. We shall provide the key steps of
this calculation below in section~\ref{sec:reparametrization}\null.

What remains is the spacetime interpretation of the variable~$\rho$. Recall
that on the initial data hypersurface $\Sigma_0$ introduced in
subsection~\ref{subsec:gauge-choice}, $\rho$ is the value of $R$ at the
sharp ridge where the hypersurface $t_1=0$ in the chart
$C_1$~(\ref{sflat1-chart}), asymptotic to $\Sigma_0$ at $r\to\infty$, meets
the hypersurface $t_2=0$ in the chart $C_2$~(\ref{sflat2-chart}), asymptotic
to $\Sigma_0$ at $r\to-\infty$.
Recall also that our Hamiltonian evolves the spacelike hypersurfaces so that
at the two infinities, covered respectively by the charts $C_1$ and~$C_2$,
we have $dt_1/dt = 1$ and $dt_2/dt=1$. One might therefore have thought
that as our initial data evolves, $\rho(t)$ would be the value of $R$ at
the sharp ridge where the hypersurface $t_1 = t$ in the chart
(\ref{sflat1-chart-left}) meets the hypersurface $t_2 = t$ in the
chart~(\ref{sflat2-chart}). However, this does not hold. The reason is that
in the $\shellpar\to0$ limit, the chart $C_1$ does not reduce to a
consistent chart across the shell, not even if one were to allow
nondifferentiability: the intrinsic metric on the shell history
is unambiguous, but evaluating this intrinsic metric from the
$\shellpar\to0$ limit of (\ref{sflat1-chart-left}) and from
(\ref{sflat1-chart-right}) leads to mutually inconsistent expressions
because the two masses differ. This means that if one approaches the shell
from the two sides on the ``same" constant
$t_1$ hypersurface, after having first taken the limit $\shellpar\to0$, one
arrives at two different two-spheres on the shell history.
The $\shellpar\to0$ limit of {\em one\/} constant $t_1$ hypersurface can be
interpreted as a continuous hypersurface in the spacetime, and this is what
we utilized in the gauge choice and the evaluation of the Liouville form,
but one cannot maintain such an interpretation for a full foliation
where~$t_1$ takes values in an open interval. The spacetime interpretation
of the variable $\rho$ must therefore be examined more carefully.

Consider the chart ${\tilde{C}}_1^-$ obtained as the $\shellpar\to0$
limit of the chart $C_1$ left of the shell. Denoting the coordinates in
${\tilde{C}}_1^-$ by $({\tilde{t}}_1,r_1)$, the metric reads
\begin{equation}
ds^2 =
-d{\tilde{t}}_1^2 + \left( dr_1 + {\sqrt{\frac{2M_-}{r_1}}} \,
d{\tilde{t}}_1\right)^2 + r_1^2d\Omega^2 ,
\quad  0 < r_1 \leq \zr
\ \ .
\label{sflat1t-chart-left}
\end{equation}
If $\tau$ is the proper time along the shell history,
we have from (\ref{sflat1-chart-right}) and (\ref{sflat1t-chart-left}) the
relation
\begin{eqnarray}
d\tau^2 &=& dt_1^2 -
\left( d\zr + {\sqrt{\frac{2M_+}{\zr}}} \,
dt_1\right)^2
\nonumber
\\
&=&
d{\tilde{t}}_1^2 - \left( d\zr + {\sqrt{\frac{2M_-}{\zr}}} \,
d{\tilde{t}}_1\right)^2
\ \ .
\label{propertimes}
\end{eqnarray}
If we fix the hypersurface ${\tilde{t}}_1 = 0$ to coincide with the
$\shellpar\to0$ limit of the initial hypersurface $t_1 = 0$ for $0 < r_1
\leq \zr$, the relation (\ref{propertimes}) determines ${\tilde{t}}_1$ as a
function of $t_1$ and the shell motion, ${\tilde{t}}_1 =
{\hat{t}}_1(t_1)$. It can now be verified that $\rho(t)$ is the value of
$R$ at the sharp ridge where the hypersurface ${\tilde{t}}_1 =
{\hat{t}}_1(t)$ in the chart (\ref{sflat1t-chart-left}) meets the
hypersurface $t_2 = t$ in the chart~(\ref{sflat2-chart}). The algebra
involved in this calculation appears not to be particularly instructive,
and it will not be reproduced here.

\subsection{Comments}

As noted above, the details of our reduction relied on certain qualitative
assumptions about the shell motion. In particular, we assumed the shell
trajectory to intersect the right-hand-side exterior region of the
Kruskal geometry right of the shell. Our gauge choice, involving the {\em
ingoing\/} spatially flat coordinates, allows us to follow the
shell trajectories into the black hole, but not into the white hole. A
time-reversed gauge choice, involving the {\em outgoing\/} spatially flat
coordinates, would conversely allow us to follow the trajectories into the
white hole but not into the black hole.

In the reduced theory~(\ref{Mminus+shell-action}), the value of the
canonical coordinate $M_-$ is a constant of motion. If we are only
interested in the shell motion, we can reduce the theory further by
dropping the Liouville term $P_- {\dot M}_-$ and regarding $M_-$ as a
prescribed positive constant. This is arguably the theory of physical
interest for an observer who scrutinizes the shell motion from one
asymptotically flat infinity and regards the ``interior" mass as
fixed. The action then reads
\begin{equation}
S = \int dt \, ( p {\dot{\zr}} - h)
\ \ ,
\label{bare-shell-action}
\end{equation}
where $h(\zr,p,M_-)$ is determined by equations
(\ref{red-shell-mom})--(\ref{h-def}).
In the limit $m\to0$, this theory reduces to that obtained by Kraus and
Wilczek \cite{kraus-wilczek1} by a less direct Hamiltonian reduction.

\section{Time-reparametrization}
\label{sec:reparametrization}

In this section we first present a general formalism for reparametrizing
time in a Hamiltonian system with a two-dimensional phase space. We then
apply this formalism to the reduced Hamiltonian
theory~(\ref{bare-shell-action}).

\subsection{General time-reparametrization formalism for two-dimensional
phase space}
\label{subsec:gen-repara}

Consider a Hamiltonian system with a two-dimensional phase space
$\boldGamma:=\{(q,p)\}$
and a time-independent Hamiltonian $h(q,p)$.
With respect to a time~$t$, Hamilton's equations read
\begin{mathletters}
\label{pa:hamilton-eqs}
\begin{eqnarray}
{dq \over dt} &=& {\partial h \over \partial p}
\ \ ,
\label{pa:hamilton-eqs-dq}
\\
{dp \over dt} &=& - {\partial h \over \partial q}
\ \ .
\end{eqnarray}
\end{mathletters}%
We wish to find a Hamiltonian system that generates the equivalent dynamics
with respect to a new parameter time~$T$, related to $t$ by
\begin{equation}
dT = N dt
\ \ ,
\label{dT=Ndt}
\end{equation}
where $N$ is a prescribed (positive) function of some suitable set of
dynamical variables. We further wish this time-reparametrization to
preserve the value of the Hamiltonian for each solution to the equations of
motion~(\ref{pa:hamilton-eqs}). We examine separately two cases: 1)~$N$ is a
function on~$\boldGamma$, and 2)~$N$ is a function of $q$ and the new
velocity $V:=dq/dT$.

\subsubsection{$N=N(q,p)$}
\label{subsubsec:Nqp}

Suppose that $N(q,p)$ is a prescribed function on~$\boldGamma$. We
replace $p$ by a new momentum $P:= {\hat P}(q,p)$, where
\begin{equation}
{\partial {\hat P}(q,p) \over \partial p} = N(q,p)
\ \ .
\label{pa:Phat-from-N}
\end{equation}
We assume that ${\hat P}(q,p)$ is an invertible function of $p$ for
each~$q$, with the inverse~${\hat p}(q,P)$. The new phase space is
$\boldGammahat:=\{(q,P)\}$, and we take the Hamiltonian on
$\boldGammahat$ to be
\begin{equation}
H(q,P) := h \biglb(q,{\hat p}(q,P)\bigrb)
\ \ .
\label{pa:Hamiltonian-def}
\end{equation}
Hamilton's equations on $\boldGammahat$ with respect to a time $T$ are
then easily seen to be equivalent to~(\ref{pa:hamilton-eqs}),
provided $t$ and $T$ are related by~(\ref{dT=Ndt}).

\subsubsection{$N=N(q,V)$}
\label{subsubsec:NqV}

Suppose next that $N(q,V)$ is a prescribed function of $q$ and the new
velocity~$V$.

Recall that equation (\ref{pa:hamilton-eqs-dq}) defines the velocity $v:=
dq/dt$ as a function on~$\boldGamma$. We
assume that this function can be inverted for the momentum as
$p = {\tilde p} (q,v)$.
We can then define on the velocity space the energy function
\begin{eqnarray}
{\tilde h}(q,v) := h\biglb(q,{\tilde p}(q,v)\bigrb)
\ \ .
\label{pa:htilde}
\end{eqnarray}
The dynamics is now encoded in the statement that ${\tilde h}(q,v)$ is
constant in~$t$. The value of ${\tilde h}(q,v)$ provides one constant of
integration, and expressing $dq/dt$ in terms of this constant and
$q$ yields the general solution in terms of a single quadrature.

Consider now the time-reparametrization (\ref{dT=Ndt}) with $N=N(q,V)$. The
velocities $v= dq/dt$ and $V= dq/dT$ are related by
\begin{equation}
v = N(q,V) V
\ \ .
\label{pa:vV}
\end{equation}
Using~(\ref{pa:vV}), we can define
on the {\em new\/} velocity space the energy function
\begin{eqnarray}
{\tilde H}(q,V) :=
{\tilde h}\biglb(q,N(q,V) V\bigrb)
\ \ .
\label{pa:Htilde}
\end{eqnarray}
Provided the relation (\ref{pa:vV}) between the velocities is not
degenerate, the full dynamics is then encoded in the statement that
${\tilde H}(q,V)$ is constant in~$T$.

We wish to find a Hamiltonian $H(q,P)$ from which ${\tilde H}(q,V)$
emerges as the energy function. If $L(q,V)$ is the corresponding
Lagrangian, we have
\begin{equation}
{\tilde H}(q,V) = V \, {\partial L (q,V) \over \partial V} - L(q,V)
\label{pa:E-from-Lag}
\end{equation}
and
\begin{equation}
P(q,V)
= {\partial L (q,V) \over \partial V}
\ \ .
\label{pa:PV}
\end{equation}
Solving (\ref{pa:E-from-Lag}) for $L(q,V)$, we find from (\ref{pa:PV}) that
the general solution for $P(q,V)$ is equivalent to
\begin{equation}
{\partial P(q,V) \over \partial V}
= V^{-1} \,
{\partial {\tilde H}(q,V) \over \partial V}
\ \ .
\label{pa:P-from-Htilde}
\end{equation}
The Hamiltonian $H(q,P)$ is obtained by inverting $P(q,V)$ for $V$ and
substituting this in ${\tilde H}(q,V)$.

\subsubsection{Comments}
\label{subsubsec:comments}

Our time-reparametrization preserves the value of the Hamiltonian on each
solution to the equations of motion. It does not, however, preserve the
value of the action, and it cannot in general be thought of as a canonical
transformation.

After $N$ is specified, the solutions to (\ref{pa:Phat-from-N}) and
(\ref{pa:P-from-Htilde}) each contain an arbitrary additive function of~$q$.
This arbitrariness corresponds to a canonical transformation that
redefines $P$ by the addition of (the gradient of) an arbitrary function.

\subsubsection{Example: relativistic particle in
(1+1)-dimensional Minkowski space}
\label{subsubsec:relPP}

As a simple example, we apply this reparametrization formalism to the free
relativistic particle in (1+1)-dimensional Minkowski space.

We start from the Hamiltonian
\begin{equation}
h= \sqrt{p^2+m^2}
\ \ ,
\label{pa:pp-h}
\end{equation}
which evolves the particle in the Minkowski
time~$t$. We then have
\begin{equation}
v=\frac{\partial h}{\partial p} = \frac{p}{\sqrt{p^2+m^2}}
\ \ .
\label{pa:pp-v}
\end{equation}
We wish to identify the new time parameter $T$ as the proper time of the
particle. {}From (\ref{dT=Ndt}) and (\ref{pa:pp-v}) we then obtain
\begin{equation}
N = \sqrt{1-v^2}
= \frac{m}{\sqrt{p^2+m^2}}
\ \ .
\label{pa:pp-N}
\end{equation}
We can thus use the above formalism with $N(q,p)=
m\left(p^2+m^2\right)^{-1/2}$. As a solution
to~(\ref{pa:Phat-from-N}), we choose ${\hat P}(q,p) = m \sinh^{-1} (p/m)$.
This leads to the familiar point particle proper time Hamiltonian
\begin{equation}
H(q,P) = m \cosh (P/m)
\ \ .
\label{pa:properH-pp}
\end{equation}

\subsection{Proper-time Hamiltonian for the self-gravitating shell}
\label{subsec:shell-proper-time}

We now apply the time-reparametrization formalism of subsection
\ref{subsec:gen-repara} to the Hamiltonian theory~(\ref{bare-shell-action}).
Our goal is to obtain a Hamiltonian that evolves the shell with respect to
its proper time. We follow the route of subsubsection~\ref{subsubsec:NqV},
specifying the reparametrization in terms of the new velocity. $M_-$ will
be regarded as a prescribed constant throughout.

We first need the Hamiltonian $h = M_+ + M_-$ (\ref{h-def}) as a function
of the old velocity~$\dot{\zr}$. Using the implicit relations
(\ref{red-shell-mom}) and (\ref{diffMmom}) to evaluate $\partial
M_+/\partial p$, we find that Hamilton's equation
${\dot{\zr}} = \partial h/\partial p$ takes the form
\begin{equation}
\dot{\zr} = {{\zp} \over \sqrt{{\zp}^2+m^2}}
- {\sqrt{\frac{2M_+}{\zr}}}
\ \ ,
\label{dot-zr-aux1}
\end{equation}
where $\zp$ is still implicitly given by~(\ref{diffMmom}). Solving
(\ref{dot-zr-aux1}) for $\zp$ and substituting in (\ref{diffMmom}) yields
\begin{equation}
{M_+ - M_- \over m}
- {m \over 2\zr}
=
{1 -
\left( \dot{\zr} + \sqrt{2M_+/\zr} \right) \sqrt{2M_+/\zr}
\over
\sqrt{
1 -
\left( \dot{\zr} + \sqrt{2M_+/\zr} \right)^2}
}
\ \ .
\label{diffMrdot}
\end{equation}
Equation (\ref{diffMrdot}) determines~$M_+$, and hence~$h$, as a function of
$\zr$ and~$\dot{\zr}$.

Let $\tau$ denote the proper time of the shell. As the parameter time $t$
coincides with the spatially flat time $t_1$ in the metric
(\ref{sflat1-chart-right}) right of the shell, we have
\begin{equation}
d\tau^2 = dt^2 -
\left( d\zr + {\sqrt{\frac{2M_+}{\zr}}} \,
dt\right)^2
\ \ .
\end{equation}
This can be solved for $dt/d\tau$ as
\begin{equation}
{dt \over d\tau}
=
{
\sqrt{2M_+ / \zr} \, (d\zr / d\tau)
+ {\tilde\epsilon}_+
\sqrt{ \left(d\zr / d\tau\right)^2
+ 1 - 2 M_+/\zr}
\over
(1 - 2M_+/\zr)
}
\ \ ,
\label{dtdtau}
\end{equation}
where the parameter ${\tilde\epsilon}_+=\pm1$ labels the two solutions.
Using (\ref{dtdtau}) to express ${\dot{\zr}}$ in terms of~$d{\zr}/d\tau$,
we can put equation (\ref{diffMrdot}) in the form
\begin{equation}
{M_+ - M_- \over m}
- {m \over 2\zr}
=
{\tilde\epsilon}_+
\sqrt{ \left(d\zr / d\tau\right)^2
+ 1 - 2 M_+/\zr}
\ \ .
\label{shell-proper-solved}
\end{equation}
As $M_+$ is a constant of motion, the shell motion is completely determined
by equation~(\ref{shell-proper-solved}). Comparing
(\ref{shell-proper-solved}) to (\ref{shell-full-eom}) shows that
our reduced Hamiltonian theory has correctly reproduced the shell motion
that arises from Israel's junction condition formalism, with the parameter
${\tilde\epsilon}_+$ coinciding with the parameter $\epsilon_+$
in~(\ref{shell-full-eom}). Equation (\ref{shell-proper-solved}) results from
squaring (\ref{shell-full-eom}) once, in a way that eliminates the
parameter~$\epsilon_-$; however, as $\epsilon_-=1$ by our global
assumptions, the full information in (\ref{shell-full-eom}) is contained
in~(\ref{shell-proper-solved}).

Solving (\ref{shell-proper-solved}) for $M_+$ yields
\begin{equation}
{M_+ - M_- \over m} + {m \over 2\zr}
=
\sqrt{ \left(d\zr / d\tau\right)^2
+ 1 - 2 M_-/\zr}
\ \ .
\label{shell-proper-solved-minus}
\end{equation}
As $M_+>M_-$, only the positive sign for the square root in
(\ref{shell-proper-solved-minus}) can occur; in terms
of~(\ref{shell-full-eom}), this sign is equal
to~$\epsilon_-$. {}From~(\ref{shell-proper-solved-minus}), the energy
function on the new velocity space reads
\begin{equation}
{\tilde{H}}(\zr,V)
=
M_+(\zr,V) + M_-
=
m
\sqrt{V^2 + 1 - 2 M_-/\zr}
- {m^2 \over 2\zr}
+ 2M_-
\ \ ,
\label{shell-proper-Htilde}
\end{equation}
where we have written, in the notation of
subsection~\ref{subsec:gen-repara},
$V:=d\zr/d\tau$. As a particular solution to (\ref{pa:P-from-Htilde}) we
choose
\begin{equation}
P(\zr,V) =
m
\ln \left( V + \sqrt{V^2 + 1 - 2 M_-/\zr} \right)
\ \ .
\end{equation}
Inverting this for $V$ and substituting in (\ref{shell-proper-Htilde})
gives the new Hamiltonian
\begin{equation}
H(\zr,P) = m \cosh (P/m) - \frac{m^2}{2\zr} + M_-\left[2-\left(
\frac{m}{\zr}\right)\exp (-P/m)\right]
\ \ .
\label{shell-proper-H}
\end{equation}

\subsection{Minkowski-like Hamiltonian for the self-gravitating shell}
\label{subsec:shell-Minkowski-time}

We now consider a time-reparametrization that makes the shell Hamiltonian
analogous to the Minkowski time point particle Hamiltonian~(\ref{pa:pp-h}),
which is also the Minkowski time Hamiltonian for a free spherical,
non-gravitating dust shell in flat space.  Starting from the shell
proper-time Hamiltonian~(\ref{shell-proper-H}), we denote the new momentum
by~$\bp$, and we run the formalism of subsubsection \ref{subsubsec:Nqp}
backwards with the choice $N(\zr,\bp)= m\left(\bp^2+m^2\right)^{-1/2}$. As
with the point particle example in subsubsection \ref{subsubsec:relPP}, we
solve (\ref{pa:Phat-from-N}) by ${\hat P}(\zr,\bp) = m \sinh^{-1} (\bp/m)$.
Denoting the counterpart of $h$ in (\ref{pa:Hamiltonian-def}) by
$\bh(\zr,\bp)$, we obtain
\begin{equation}
\bh(\zr,\bp) =
\sqrt{\bp^2+m^2}
-{m^2 \over 2\zr}
+ M_-
\left[ 2 - {1 \over \zr}
\left( \sqrt{\bp^2+m^2}-\bp \right)
\right]
\ \ .
\label{bh}
\end{equation}

\section{$\RPthree$ geon with a self-gravitating shell}
\label{sec:geon}

In this section we adapt the formalism to a shell in a spacetime
with the $\RPthree$ geon topology.

As mentioned in the Introduction, an asymptotically flat, spherically
symmetric spacetime $(M, g)$ with a single asymptopia can have spatial
topology $\RPthree \setminus${}$\{$a point at infinity$\}$. We refer to an
asymptotically flat spacetime with this topology, or to an asymptotically
flat initial data set in such a spacetime, as an $\RPthree$ geon. The
covering space of the spacetime then has the wormhole topology of the
extended Schwarzschild geometry.

In vacuum, one can obtain a spherically symmetric $\RPthree$ geon Einstein
spacetime as the quotient of Kruskal manifold under a freely and
properly discontinuously acting involutive isometry \cite{topocen}. Let
$(\bar M, \bar g)$ be Kruskal manifold, and let $({\tilde t},{\tilde
x},\theta,\phi)$ be a chart in which ${\tilde t}$ and ${\tilde x}$ are the
usual Kruskal time and space coordinates (denoted respectively by $v$ and
$u$ in Ref.\ \cite{MTW}).  The isometry in question is then
\begin{equation}
I:({\tilde t},{\tilde x},\theta,\phi)
\mapsto
({\tilde t},-{\tilde x}, \pi - \theta,\phi + \pi)
\ \ .
\end{equation}
As $I$ commutes with rotations, the quotient spacetime
$(M, g):=(\bar M, \bar g)/I$ is spherically symmetric. In $(\bar M, \bar
g)$, the constant ${\tilde t}$ hypersurfaces that do not hit a singularity
have topology $S^2\times\BbbR$, with two asymptotically flat
infinities, and they have at ${\tilde x}=0$ a wormhole throat at which the
radius of the
$S^2$ reaches its minimum value. In $(M, g)$, the corresponding constant
${\tilde t}$ hypersurfaces have topology $\RPthree \setminus${}$\{$a
point at infinity$\}$, and the throat has become a ``minimum radius"
two-surface with topology~$\RPtwo$. Away from the throat history,
$(M,g)$ is indistinguishable from half (say, ${\tilde x}>0$) of $(\bar M,
\bar g)$. The Penrose diagram can be found in Ref.\ \cite{topocen}. Note
that the throat history in $(\bar M, \bar g)$ is only defined with
respect to a given foliation,
while the throat history in $(M,g)$ has a
coordinate invariant meaning as the trajectory of the ``minimum
radius"~$\RPtwo$. The reason for this difference is that $I$ does not
commute with the Killing time translations on
$(\bar M,\bar g)$: these Killing time translations do not descend into
globally-defined isometries of $(M,g)$.

Consider now a spherically symmetric spacetime that has the $\RPthree$ geon
topology and solves Einstein's equations with a spherical dust shell. Away
from the shell, Birkhoff's theorem still guarantees that the spacetime is
locally isometric to Kruskal manifold. We assume that the spacetime right of
the shell is as in  section \ref{sec:reduction}: this part of the spacetime
is part of Kruskal geometry, containing the right-hand-side Kruskal
infinity, and the shell trajectory intersects the right-hand-side exterior
region in this Kruskal geometry. The spacetime left of the shell is assumed
to be part of the vacuum $\RPthree$ geon spacetime described above, and to
contain the throat history.

If the shell passes through the throat, it needs to cross itself there. We
assume that such a crossing does not happen.

A Cauchy surface in this spacetime has only one infinity, in the part right
of the shell, whereas the part
left of the shell is compact. We can therefore unambiguously regard the
left-hand side of the shell as the interior and the right-hand side as the
exterior.

It is easy to adapt the Hamiltonian formalism of section \ref{sec:action} to
these $\RPthree$ boundary conditions. We take $N$, $\Lambda$, $R$,
$\pi_\Lambda$, and $\pi_R$ to be even in~$r$ and $N^r$ odd in~$r$, with
the consequence that $M_+=M_-$ in the falloff~(\ref{l-r}). We assume
$\zr>0$, add to the system a second shell at $r=-\zr$, and finally take the
quotient of the spacetime under the isometry $(t,r,\theta,\phi)\mapsto
(t,-r,\pi - \theta,\phi+\pi)$. The resulting Hamiltonian theory is clearly
consistent in the same sense as the Kruskal-type theory of
section~\ref{sec:action}\null.
The action can be written as
\begin{equation}
S = S_\Sigma + S_{\partial\Sigma}
\ \ ,
\label{total-action}
\end{equation}
where $S_\Sigma$ is given by (\ref{bulk-action}), with the $r$-integration
extending from $r=0$ to $r=\infty$, and
\begin{equation}
S_{\partial\Sigma} = - \int dt \, M_+
\ \ .
\label{geon:boundary-action}
\end{equation}
When the equations of motion hold, we recover the above $\RPthree$ Einstein
spacetimes with a dust shell. The throat is located at $r=0$.

The Hamiltonian reduction proceeds in close analogy with that
in section~\ref{sec:reduction}. To choose the gauge, we introduce the
analogue of the chart $C_1$~(\ref{sflat1-chart}), with $M_->0$ now denoting
the mass in the interior. The range of $r_1$ is bounded below by the
$t_1$-dependent throat radius, and it is the throat radius that emerges as
the parameter~$\rho$. The transition region near
the throat is handled as in appendix~\ref{app:ridge-trans}, but because
now $r>0$ in our action, the contribution to the Liouville form is only half
of that found in appendix~\ref{app:ridge-trans}. The transition region near
the shell is handled exactly as in appendix~\ref{app:shell-trans}. The
reduced action is given by equations
(\ref{red-moms})--(\ref{red-action-rho}), with the exceptions that the
right-hand side in the counterpart of (\ref{rho-mom}) contains the
factor~$\casehalf$, and (\ref{h-def}) is replaced by
\begin{equation}
h := M_+
\ \ .
\label{geon:h-def}
\end{equation}
{}From section \ref{sec:reduction} it is clear that the reduced theory
reproduces the correct equations of motion. In the classical
solutions, $\rho(t)$ is the value of $R$ at the throat in a foliation
defined as with the chart~(\ref{sflat1t-chart-left}).

A canonical transformation that replaces the pair $(\rho,p_\rho)$ by
$(M_-,P_-)$ leads to the action~(\ref{Mminus+shell-action}), with
(\ref{h-def}) replaced by~(\ref{geon:h-def}). Dropping the term $P_-{\dot
M}_-$ gives a theory in which the interior mass $M_-$ is regarded as a
prescribed positive
constant. The time-reparametrizations of section
\ref{sec:reparametrization} clearly carry through without change: the
counterparts of the Hamiltonians (\ref{shell-proper-H}) and (\ref{bh})
differ only in that the (constant) additive term $2M_-$ is replaced
by~$M_-$.

\section{Self-gravitating shell with $\BbbR^3$ spatial topology}
\label{sec:flat-interior}

In this section we consider the spatial topology~$\BbbR^3$.

We start directly from the action principle. In the bulk
action~(\ref{bulk-action}), we take $0<r<\infty$, with the falloff
(\ref{l-r}) at $r\to\infty$. The total action is given by
(\ref{total-action}) and~(\ref{geon:boundary-action}). At $r\to0$, we
introduce the falloff
\begin{mathletters}
\label{s-metricfall}
\begin{eqnarray}
\Lambda(t,r)
&=&
\Lambda_0 + O(r^2)
\ \ ,
\label{s-Lambdafall}
\\
R(t,r)
&=&
R_1 r + O(r^3)
\ \ ,
\label{s-Rfall}
\\
\pi_\Lambda(t,r)
&=&
\pi_{\Lambda_2} r^2 + O(r^4)
\ \ ,
\label{s-PLambdafall}
\\
\pi_R(t,r) &=&
\pi_{R_1} r + O(r^3)
\ \ ,
\label{s-PRfall}
\\
N(t,r)
&=&
N_0 +
O(r^2)
\ \ ,
\label{s-Nfall}
\\
N^r(t,r)
&=&
N^r_1 r + O(r^3)
\ \ ,
\label{s-Nrfall}
\end{eqnarray}
\end{mathletters}%
where $\Lambda_0>0$, $R_1>0$, $\pi_{\Lambda_2}$, $\pi_{R_1}$, $N_0>0$, and
$N^r_1$ are functions of $t$ only. It is straightforward to verify that
this falloff is consistent with the constraints
and preserved by the time evolution, and no
additional boundary terms in the action are needed at $r=0$. {}From
(\ref{pi-sol}) we see that in the classical solutions, the mass left of the
shell must vanish, and $r=0$ is just the coordinate singularity at the
center of hyperspherical coordinates in flat space. The classical solutions
therefore describe a self-gravitating shell with a flat interior.
The spatial topology is~$\BbbR^3$.

The reduction proceeds as above, using the analogue of the chart
$C_1$~(\ref{sflat1-chart}) with $M_-=0$ and $r_1>0$. In the region
$r_1<\zr-\shellpar$, the initial data hypersurface $\Sigma_0$ extends
smoothly to $r_1=0$, and there is no counterpart of the parameter $\rho$ of
the Kruskal and $\RPthree$-geon topologies.
The only contribution to the Liouville
form comes from the shell transition region, which is handled exactly as in
appendix~\ref{app:shell-trans} but with $M_-=0$. The reduced action is
given by (\ref{bare-shell-action}) and~(\ref{geon:h-def}), where $M_+$ is
obtained from (\ref{red-shell-mom}) and (\ref{diffMmom}) with $M_-=0$. It is
again clear from section \ref{sec:reduction} that this reduced
theory reproduces the correct dynamics.
As expected, the reduced phase space is two-dimensional.

The time-reparametrizations of section \ref{sec:reparametrization}
carry through without change. The counterparts of the Hamiltonians
(\ref{shell-proper-H}) and (\ref{bh}) are obtained
from these formulas
by simply setting $M_-=0$. In particular, (\ref{bh}) reduces to the
Hamiltonian used in Refs.\ \cite{HKK,dolgov}.

\section{Remarks on quantization}
\label{sec:quantization}

In this section we
discuss the prospects for quantizing the reduced
theories. We first review the pure vacuum case, and then turn to the
coupled system.

\subsection{Mass spectrum of spherically symmetric vacuum wormholes and
$\RPthree$ geons}
\label{subsec:vacuum-quantization}

In sections \ref{sec:action}--\ref{sec:flat-interior} we considered the
dynamics of a shell coupled to
spacetime geometry. However, the methods immediately adapt to spherically
symmetric vacuum gravity by simply omitting the shell.

With the Kruskal topology, and no asymptotic masses fixed,
the constraints imply $M_+=M_-:=M$. With the gauge choice of
section~\ref{sec:reduction}, without the shell, the reduced action
reads
\begin{equation}
S = \int dt \, ( p_\rho {\dot\rho} - h)
\ \ ,
\end{equation}
where $h=2M$, and $M$ is obtained from (\ref{rho-mom}) with $M_-=M$.
Geometrically, $\rho(t)$ is the value of $R$ at the sharp ridge in the
foliation described in subsection \ref{subsec:gauge-choice}, without the
shell; this ridge evolves in the black hole interior along a history of
constant Killing time. At $t\to-\infty$, we have $\rho(t)\to2M$ as the ridge
approaches the bifurcation two-sphere, but in the future the gauge breaks
down at a finite value of $t$ as $\rho(t)\to0$. With the $\RPthree$-geon
topology, the only differences are that $h=M$, and the right-hand
side of (\ref{rho-mom}) contains an additional factor~$\casehalf$. The
reduced phase space is two-dimensional in each case. These results agree
with those obtained by Kucha\v{r}'s reduction method \cite{kuchar1,LW2}
under a falloff that is qualitatively similar but makes the constant $t$
hypersurfaces asymptotic to hypersurfaces of constant Minkowski time.

If one chooses to fix the mass at one infinity with the Kruskal topology,
the reduced theory has no degrees of freedom. The same holds if one chooses
to fix the mass at the infinity or at the throat with the $\RPthree$ geon
topology. With the $\BbbR^3$ topology, the reduced theory is always void.

Quantizing the reduced theories with a zero-dimensional reduced phase space
is of course trivial: the mass $M$ is a prescribed $c$-number. Quantizing
the theories with a two-dimensional reduced phases space offers, however,
several options.

One option is to perform first a canonical transformation to the
pair $(M,P_M)$ as in sections \ref{sec:reduction} and~\ref{sec:geon}\null.
One can then take quantum states to be described by functions $\Psi(M)$ of
the positive-valued configuration variable~$M$, adopt the inner product
$\langle\Psi_1 | \Psi_2 \rangle = \int_0^\infty dM \, \overline{\Psi_1(M)}
\Psi_2(M)$ (or a similar inner product with some $M$-dependent weight
factor), and promote $M$ into the quantum operator $\hat{M}$ that acts in
the Schr\"odinger picture as \cite{thiemann1,thiemann2,kuchar1}
\begin{equation}
\hat M \Psi (M) = M \Psi (M)
\ \ .
\label{Mhat-oper}
\end{equation}
The spectrum of $\hat{M}$, and thus also that of the Hamiltonian
operator~$\hat{h}$, is continuous and consists of the positive real axis.

Another option is to take quantum states to be described by
functions $\psi(\rho)$ of the positive-valued ``throat radius"~$\rho$,
adopt the inner product $\langle\psi_1 | \psi_2 \rangle = \int_0^\infty
\mu(\rho) d\rho \,
\overline{\psi_1(\rho)}
\psi_2(\rho)$ where $\mu(\rho)$ is some weight factor, and try to promote
the function $M(\rho,p_\rho)$
into an operator on this Hilbert space. As our $M(\rho,p_\rho)$
is known only implicitly, we have not tried to pursue
this quantization, but there seems no obvious reason to expect that the
spectral properties of the resulting Hamiltonian operator would agree with
those of the operator $\hat{M}$ in~(\ref{Mhat-oper}).

Indeed, quantization of spherically symmetric vacuum gravity was discussed
in Ref.\ \cite{lm} in terms of a related ``wormhole throat" phase space
$(a,p_a)$, on which the Schwarzschild mass
is given by
\begin{equation}
M(a,p_a)=\frac{1}{2}
\left( \frac{p_a^2}{a} + a\right)
\ \ .
\label{Mofap}
\end{equation}
The configuration variable $a$ has an interpretation as the radius of the
wormhole throat, much as our~$\rho$, but with a time parameter that is now
identified with the proper time of the throat history.\footnote{The
Hamiltonian $M(a,p_a)$ (\ref{Mofap}) describing the proper-time evolution of
the throat was previously considered by Friedman, Redmount and
Winters-Hilt \cite{fried-red-win,redmount-talk} without a derivation by
reduction from spherically symmetric vacuum gravity.
In Ref.\ \cite{lm}, this Hamiltonian was derived from Kucha\v{r}'s reduced
Hamiltonian theory \cite{kuchar1} by a canonical transformation. A similar
derivation could clearly be given from the canonical pair $(M,P_M)$ of the
present paper, despite the technical differences in our falloff and that of
Ref.\ \cite{kuchar1}.}
If the Hilbert space is chosen as above with the configuration
variable~$\rho$, with reasonable choices for the weight factor~$\mu$,
the function $M(a,p_a)$ can be promoted into a self-adjoint operator whose
spectrum is bounded below and purely discrete \cite{lm}.

We regard as artificial the continuous mass spectrum arising from the
quantization~(\ref{Mhat-oper}), because one can similarly obtain a
continuous spectrum for {\em any\/} dynamical system whose Hamiltonian
is not explicitly time-dependent.  For any function $H$ with
nonvanishing gradient on the phase space, one can find a local
canonical chart of the form
$(H, q_2, \cdots , q_n, p_H, p_2, \cdots , p_n)$,
in which $H$ is one of the
canonical coordinates. If the range of $H$ in this chart is~$\BbbR_+$, one
can adopt a Schr\"{o}dinger representation with Hilbert space $L_2(\BbbR_+)
\otimes \sfH$, with $\sfH$ a Hilbert space for the remaining~$q$'s. The
Hamiltonian operator $\hat H$ can then be taken to act as a multiplication
operator,
\begin{equation}
\hat H \psi (H, q_2,\cdots ,
q_n) = H\psi (H, q_2, \cdots , q_n)
\ \ ,
\end{equation}
and its spectrum is~$\BbbR_+$.

Ambiguities in canonical quantization are, of course, well recognized
\cite{woodhouse,isham-LH,AAbook}. One specific issue not addressed above is
in the global properties of the canonical transformations. For example, the
canonical transformation that takes the phase space $(a,p_a)$ to
Kucha\v{r}'s reduced phase space \cite{kuchar1}
is not onto: the classical dynamics
in Kucha\v{r}'s reduced phase space
is complete, but the classical dynamics in the
phase space $(a,p_a)$ is not \cite{lm}. One's attitude to such classical
incompleteness in view of quantization may depend on what one sees as the
role of  singularities in quantum gravity
\cite{lm,gotay-dem,poisson-israel,bonanno,marolf-recoll,%
horo-marolf,horo-myers,droz}.

\subsection{Quantization of shell coupled to geometry}

We now turn to the coupled system. We restrict consideration to the
proper-time Hamiltonian~(\ref{shell-proper-H}).

When $M_-=0$, the shell encloses a flat interior with trivial topology,
and the Hamiltonian (\ref{shell-proper-H}) takes the form corresponding to
a relativistic particle in a Coulomb potential,
\begin{equation}
H(\zr,P) = m \cosh (P/m) - \frac{m^2}{2\zr}
\ \ ,
\label{shell-proper-H-flat}
\end{equation}
discussed by H\'aj\'{\i}\v{c}ek \cite{hajicek}. One can adopt a
Schr\"{o}dinger representation corresponding to configuration-space
variable $\zr \in {\Bbb R}_+$ and Hilbert space
\begin{equation}
\sfH
:=
L_2(\BbbR_+, r^\alpha d\zr )
\ \ ,
\end{equation}
where $\alpha$ is a parameter.
With the factor ordering
\begin{equation}
\widehat{[\cosh (P/m)]}
= \lim_{N\rightarrow\infty}
\zr^{-\alpha/2} \sum_{n=0}^N \frac{(-1)^n}{(2n)!} \Delta^n
\zr^{\alpha/2}
\ \ ,
\label{cosh-operator}
\end{equation}
where $\Delta = \partial^2_{\zr}$, $H$ becomes a self-adjoint operator
$\hat H$ with domain \cite{ow}
\begin{equation}
D(\hat H) = \left\{ f \mid f^{(2n)}(0) = 0,\,
f^{(n)}\in L_2,\,
\hbox{all $n$}\right\}
\ \ .
\end{equation}
(H\'aj\'{\i}\v{c}ek takes $\alpha =-1$, but notes the unitary
equivalence of $(\sfH, \hat{H})$ for a different choice of~$\alpha$.)  For
$m < 1.9$, $\hat{H}$ is bounded below, and its spectrum, like that
of the nonrelativistic Coulomb problem, has discrete and continuous
parts.

When $M_->0$, one expects that the Hamiltonian (\ref{shell-proper-H}) can
be made into a self-adjoint operator in an analogous manner, and one
expects the spectrum then to be bounded below and partly discrete for small
values of~$m$. However, there appears to be no reason to expect that the
term proportional to $M_-$ would allow the spectrum to have a
lower bound for large values of~$m$.  Oharu and Winters-Hilt
\cite{ow} are currently examining a self-adjoint extension of $H$ on
$L_2(\BbbR_+, d\zr)$, with factor ordering corresponding to the choice
(\ref{cosh-operator}) with
$\alpha =0$:
\begin{equation}
\hat H = m \widehat{[\cosh (P/m)]}
- \frac{m^2}{2\zr} -
M_- m
\zr^{-1/2}
\widehat{[\exp (-P/m)]}
\zr^{-1/2}
+ 2 M_-
\ \ .
\end{equation}

Finally, recall that our time-reparametrization derivation of the
proper-time Hamiltonian (\ref{shell-proper-H}) assumed $M_-$ to be a
prescribed, time-independent constant.
With $\BbbR^3$ spatial topology this assumption is automatically satisfied.
With the Kruskal and $\RPthree$-geon topologies, on the other hand,
one could ask whether it is still possible to carry out an analogous
time-reparametrization when $M_-$ is a dynamical variable and the phase
space is four-dimensional.
If the answer is affirmative, one could presumably
raise anew the issues regarding the spectrum of ${\hat{M}}_-$ that were
addressed in the context of the vacuum theory in
subsection~\ref{subsec:vacuum-quantization}\null.
If, after the reparametrization,
the dynamics of $M_-$ still decouples from the dynamics of the shell as in
sections \ref{sec:reduction} and~\ref{sec:geon}, one could effectively
separate variables by first considering the eigenvalue equation
for~${\hat{M}}_-$,
\begin{equation}
\hat M_-\psi = M_-\psi
\ \ .
\end{equation}
For each eigenspace of~$\hat M_-$, the shell Hamiltonian would then have the
form (\ref{shell-proper-H}) with a $c$-number~$M_-$, and the character of
the total spectrum would depend on the spectrum of~${\hat{M}}_-$.

\section{Summary and discussion}
\label{sec:discussion}

In this paper we have considered the Hamiltonian dynamics of spherically
symmetric spacetimes that contain an idealized, infinitesimally thin
massive dust shell. We considered the Kruskal-like
spatial topology $S^2\times\BbbR$, the $\RPthree$-geon spatial topology
$\RPthree\setminus${}$\{$a point at infinity$\}$, and the Euclidean spatial
topology~$\BbbR^3$. The variational equations that arose
from the unreduced Hamiltonian action were not
strictly consistent in a distributional sense, but we were able to localize
the ambiguity into the single equation that arises by varying the action
with respect to the shell position. When the ambiguous contribution to this
equation was interpreted as the average of its values on the two sides of
the shell, we correctly reproduced the content of Israel's junction
condition formalism.

We performed a Hamiltonian reduction by adopting a gauge with piecewise flat
spatial sections, and passing to the limit in which the interpolating
transition regions became vanishingly narrow. The constraints could then
be explicitly solved. For the Kruskal and $\RPthree$ topologies the reduced
phase space was four-dimensional, with one canonical pair closely associated
with the shell motion and the other pair with the dynamics of the geometry.
In the limit where the shell is not present, this correctly reproduced
previous results for spherically symmetric vacuum geometries. Retaining the
shell but prescribing by hand one asymptotic mass for the Kruskal topology,
and the interior mass for the $\RPthree$ topology, we recovered theories
whose reduced phase space was two-dimensional, with just the canonical pair
associated with the shell motion surviving. For the $\BbbR^3$ topology, the
interior mass necessarily vanishes, and we only obtained a two-dimensional
phase space, with the single canonical pair describing the shell motion.

For each of the three spatial topologies, we time-reparametrized the
dynamics in the two-dimensional phase space that describes the shell motion
with fixed interior mass. With one choice for the reparametrization, we
recovered a previously known Hamiltonian that evolves the shell with
respect to its proper time. With another choice, we recovered a Hamiltonian
analogous to the square-root Hamiltonian of a spherical test shell in
Minkowski space. Finally, we briefly discussed the spectra that would be
expected to emerge in different approaches of canonically quantizing the
theories.

Our results provide a robust description of the reduced Hamiltonian dynamics
of a spherically symmetric dust shell coupled to gravity, in the region of
the reduced phase space that is covered by our piecewise spatially flat
gauge.
While this gauge is not global, one can argue that this gauge and its
time-inverted counterpart cover the region of the reduced phase space that
is of interest to an observer who scrutinizes the shell motion from one
asymptotically flat infinity.
What remains open, however, is the global structure of the reduced phase
space. One would also like to describe the reduced phase space in a way
that is more geometrical and less tied to a particular gauge. One possible
avenue for this, currently under investigation by H\'aj\'{\i}\v{c}ek and
Kijowski \cite{haji-kijo1,haji-kijo2},
might be to generalize to the massive
shell the canonical transformations that Kucha\v{r} introduced to simplify
the vacuum theory \cite{kuchar1}. Work on the analogous problem with a
null-dust shell is in progress \cite{fri-lou-whit-null}.

More ambitiously, one would like to consider systems with matter that is
more interesting than a dust shell. The canonical formulation of Einstein
gravity coupled to a {\em continuous\/} distribution of massive or null dust
has been discussed respectively in Refs.\ \cite{B+K} and
\cite{bicak-kuchar-ndust}. For the canonical formulation in the presence of
other types of fluids, see for example Ref.\ \cite{brown-marolf} and the
references therein. A discussion of the difficulties involved with
spherically symmetric gravity coupled to a scalar field is given in Refs.\
\cite{romano1,romano2}. A discussion in the context of a dilatonic black
hole can be found in Ref.\ \cite{KRV}.

\acknowledgments
We would like to thank
Paul Branoff,
Valery Frolov,
Petr H\'aj\'{\i}\v{c}ek,
Jerzy Kijowski,
Karel Kucha\v{r},
Jarmo M\"akel\"a,
Charlie Misner,
Eric Poisson,
Ian Redmount,
Rafael Sorkin,
and Bernard Whiting
for helpful discussions.
This work was supported in part by NSF grants
PHY-94-21849
and PHY-95-07740.

\appendix
\section{Ridge transition region}
\label{app:ridge-trans}

In this appendix we specify the gauge in the ridge transition region
$|r| \le (1+\ridgepar)\rho$, and evaluate the contribution from this
region to the integral on the right-hand side of (\ref{theta}) in the limit
$\ridgepar\to0$.

\subsection{Gauge choice}
\label{subapp:ridge-gauge}

To specify the gauge in the region $|r| \le (1+\ridgepar)\rho$, we consider
the classical spacetime of subsection~\ref{subsec:gauge-choice}, and the
spacelike hypersurface
${\tilde\Sigma}_0$ in this spacetime. The part $|r| \le (1+\ridgepar)\rho$
of ${\tilde\Sigma}_0$ lies in the
black-hole region of the Kruskal spacetime left of the shell.

Let $h:\BbbR\to\BbbR$ be a smooth function such that
\begin{equation}
h(x) = \left\{
\begin{array}{l}
0 \\ x - \casehalf
\end{array}
\begin{array}{l} , \hspace{2mm}
x \le 0 ,\\ , \hspace{2mm}
x \ge 1 , \end{array}
\right.
\label{hfunc-def}
\end{equation}
and $d^2 h/dx^2 >0$ for $0<x<1$. We write $h^{(n)}(x) := d^n h(x)/dx^n$.

For $|r| \le (1+\ridgepar)\rho$, we seek a
gauge in the form
\begin{mathletters}
\label{ridgepar-gauge}
\begin{eqnarray}
\Lambda(r) &=& (1 - \Lambda_0) h^{(1)}\!\left({|r| - \rho \over
\ridgepar\rho}\right) + \Lambda_0
\ \ ,
\label{ridgepar-gauge-Lambda}
\\
R(r) &=& \left[ \ridgepar h\!\left({|r| - \rho \over
\ridgepar\rho}\right) + 1 + \casehalf\ridgepar
\right]
\rho
\ \ ,
\label{ridgepar-gauge-R}
\end{eqnarray}
\end{mathletters}%
where $\Lambda_0$ is a positive parameter. In the subregion $|r| \le \rho$,
the radius of the two-sphere is constant on~${\tilde\Sigma}_0$, $R = (1 +
\casehalf\ridgepar)\rho$, and the proper distance on ${\tilde\Sigma}_0$
is~$\Lambda_0 dr$. The subregions $\rho \le |r| \le (1+\ridgepar)\rho$
interpolate smoothly between this constant radius gauge and the
spatially flat gauge~(\ref{simpleflatgauge}). Note that $\Lambda(r)>0$, and
$(1 + \casehalf\ridgepar)\rho \le R \le (1 +\ridgepar)\rho$.

Recall from subsection \ref{subsec:gauge-choice} that $(1+\ridgepar)\rho <
2M_-$. Equations (\ref{pi-sol}) thus yield a real-valued solution for
$\pi_\Lambda$ and $\pi_R$ for all of $|r| \le (1+\ridgepar)\rho$. The gauge
(\ref{ridgepar-gauge}) therefore specifies a spacelike hypersurface in an
interior Kruskal geometry with mass~$M_-$, with the ends at $R =
(1+\ridgepar)\rho$. What remains is to choose the parameter $\Lambda_0$ in
(\ref{ridgepar-gauge-Lambda}) so that this hypersurface precisely fits
between the points $|r| = (1+\ridgepar)\rho$.

In the curvature coordinates $(T,R)$ in the black hole interior,
the metric reads
\begin{equation}
ds^2 =
- \left({2M_-\over R} - 1\right)^{-1} dR^2
+ \left({2M_-\over R} - 1\right) dT^2
+ R^2 d\Omega^2
\ \ ,
\label{int-curv-metric}
\end{equation}
where $0<R<2M_-$, $R$ decreases to the future, and we take $T$ to increase
to the right. The transformation
from (\ref{int-curv-metric}) to the chart $C_1$ of
subsection~\ref{subsec:gauge-choice} reads
\begin{mathletters}
\label{curvc1trans}
\begin{eqnarray}
T &=&
t_1 - 2 \sqrt{2 M_- r_1}
- 2 M_- \ln
\left(
\sqrt{2 M_-} - \sqrt{r_1}
\over
\sqrt{2 M_-} + \sqrt{r_1}
\right)
\ +\hbox{constant}
\ \ ,
\\
R &=& r_1
\ \ .
\end{eqnarray}
\end{mathletters}%
As our (prospective) deformation (\ref{ridgepar-gauge}) of ${\hat\Sigma}_0$
to ${\tilde\Sigma}_0$ is symmetric around $r=0$, the value of $T$ at $r=0$
on ${\tilde\Sigma}_0$ is the same as the value of $T$ at the unsmoothed
ridge on~${\hat\Sigma}_0$. On ${\tilde\Sigma}_0$, we thus have
\begin{eqnarray}
T_{r = (1+\ridgepar)\rho} - T_{r = 0}
&=&
2 \sqrt{2 M_- \rho}
- 2 \sqrt{2 M_- (1+\ridgepar) \rho}
\nonumber
\\
&&
+ 2 M_- \ln
\left[
(\sqrt{2 M_-} - \sqrt{\rho})
\left(\sqrt{2 M_-} + \sqrt{(1+\ridgepar)\rho}\right)
\over
(\sqrt{2 M_-} + \sqrt{\rho})
\left(\sqrt{2 M_-} - \sqrt{(1+\ridgepar)\rho}\right)
\right]
\ \ .
\label{Tdiff}
\end{eqnarray}
On the other hand, from Eq.~(80) of Ref.\
\cite{kuchar1} we have
\begin{equation}
T' = {\Lambda \pi_\Lambda \over 2M_- - R}
\ \ .
\label{Tprime}
\end{equation}
Integrating (\ref{Tprime}) from $r=0$ to $r=(1+\ridgepar)\rho$, with
$\pi_\Lambda$ given by~(\ref{pi-lambda-sol}), and equating the result
to~(\ref{Tdiff}), gives a relation that implicitly determines $\Lambda_0$
in terms of~$M_-$, $\rho$, and~$\ridgepar$. By the symmetry of
${\tilde\Sigma}_0$ around $r=0$, the relation obtained by similarly
comparing $T_{r = - (1+\ridgepar)\rho}$ to $T_{r = 0}$, using the
chart~$C_2$, contains exactly the same information. This completes the
gauge choice.

We shall below be interested in the limit of small~$\ridgepar$. In this
limit, the relation determining $\Lambda_0$ admits a
power series expansion in~$\ridgepar$. The result is
\begin{equation}
\Lambda_0 = {\ridgepar \over 2 \sqrt{1 - \rho/(2M)}} + O(\ridgepar^2)
\ \ ,
\label{Lambdanought-expansion}
\end{equation}
where $O$ stands for a $\ridgepar$-dependent quantity that is
bounded by a constant times its argument.

\subsection{Liouville form}
\label{subapp:ridge-Liouville}

We now evaluate the contribution to the integral in the Liouville form
(\ref{theta}) from $|r| \le (1+\ridgepar)\rho$, in the limit
$\ridgepar\to0$.

As we have noted, the gauge (\ref{ridgepar-gauge}) for $|r| \le
(1+\ridgepar)\rho$ joins smoothly to the spatially flat gauge outside this
interval, and the expressions given in (\ref{ridgepar-gauge}) are in fact
valid for all of $-\infty < r < \zr - \shellpar$. The differentials $\extder
\Lambda(r)$ and $\extder R(r)$ therefore contain no delta-functions in $r$
at $|r| = (1+\ridgepar)\rho$, and it is sufficient to consider the
contributions from $|r| < \rho$ and $\rho < |r| < (1+\ridgepar)\rho$.

For $|r| < \rho$, we have $h = h^{(1)}=0$. Equations
(\ref{ridgepar-gauge}) and (\ref{Lambdanought-expansion}) yield
$\extder\Lambda = O(\ridgepar)$ and $\extder R = O(1)$, and equations
(\ref{pi-sol}) yield $\pi_\Lambda = O(1)$ and $\pi_R
= O(\ridgepar)$. The contribution to (\ref{theta})
is therefore~$O(\ridgepar)$.

Suppose then $\rho < r < (1+\ridgepar)\rho$. We now obtain
\begin{mathletters}
\begin{eqnarray}
\extder \Lambda
&=&
- {h^{(2)} \over \ridgepar \rho}  \, \extder \rho
+ O(1)
\ \ ,
\\
\extder R
&=&
\left( 1 - h^{(1)} \right) \extder \rho
+ O(\ridgepar)
\ \ ,
\\
\pi_\Lambda
&=&
\rho \sqrt{(R'/\Lambda)^2-1+2M_-/R}
+ O(\ridgepar)
\ \ ,
\\
\pi_R
&=&
{ \rho (R'/\Lambda)'
\over
\sqrt{(R'/\Lambda)^2-1+2M_-/R} }
+ O(1)
\ \ ,
\label{piRridgefall}
\end{eqnarray}
\end{mathletters}%
where the argument of $h$ and its derivatives is always
$(r-\rho)/\ridgepar\rho$. Note that the first term in (\ref{piRridgefall})
is~$O(\ridgepar^{-1})$.

For $\int dr \, \pi_\Lambda \extder\Lambda$,
changing the integration variable from $r$ to $x:=(r -
\rho)/\ridgepar\rho$ gives
\begin{eqnarray}
\int_\rho^{(1+\ridgepar)\rho} dr \, \pi_\Lambda \extder\Lambda
&=&
- \rho \extder\rho \int_0^1 dx \, h^{(2)}(x)
\sqrt{(R'/\Lambda)^2-1+2M_-/R}
\ + O(\ridgepar)
\nonumber
\\
&=&
- \sqrt{2M_-\rho} \, \extder\rho
\int_0^1 dx \, h^{(2)}(x)
\ + o(1)
\nonumber
\\
&=&
- \sqrt{2M_-\rho} \, \extder\rho
+ o(1)
\ \ ,
\end{eqnarray}
where $o(1)$ stands for a $\ridgepar$-dependent quantity that goes to zero
as $\ridgepar\to0$. We have used the fact that
$\sqrt{(R'/\Lambda)^2-1+2M_-/R}\to \sqrt{2M_-/\rho}$ pointwise in $x$
as $\ridgepar\to0$, and taken the limit under the integral by dominated
convergence.

For $\int dr \, \pi_R \extder R$, the assumption $h^{(2)}>0$ allows us to
change the integration variable from $r$ to $u:=R'/\Lambda$. We obtain
\begin{eqnarray}
\int_\rho^{(1+\ridgepar)\rho} dr \, \pi_R \extder R
&=&
\rho \, \extder \rho
\int_\rho^{(1+\ridgepar)\rho}
{ dr \, (R'/\Lambda)' \left( 1 - h^{(1)} \right)
\over
\sqrt{(R'/\Lambda)^2-1+2M_-/R} }
\ + O(\ridgepar)
\nonumber
\\
&=&
\rho \, \extder \rho
\int_0^1
{ du \left( 1 - h^{(1)} \right)
\over
\sqrt{(R'/\Lambda)^2-1+2M_-/R} }
\ + O(\ridgepar)
\nonumber
\\
&=&
\rho \, \extder \rho
\int_0^1
{ du
\over
\sqrt{u^2-1+2M_-/\rho} }
\ + o(1)
\nonumber
\\
&=&
\casehalf
\ln
\left(
\sqrt{2 M_-} + \sqrt{\rho}
\over
\sqrt{2 M_-} - \sqrt{\rho}
\right)
\rho \, \extder \rho
+ o(1)
\ \ .
\end{eqnarray}
We have used the facts that $\sqrt{(R'/\Lambda)^2-1+2M_-/R}\to
\sqrt{u^2-1+2M_-/\rho}$ and $h^{(1)} = u \Lambda_0 \left[ 1 - u(1-\Lambda_0)
\right]^{-1}\to 0$ pointwise in $u$ as
$\ridgepar\to0$, and taken the limit under the integral by dominated
convergence.

Adding the identical contributions from the region $-(1+\ridgepar)\rho < r
< -\rho$, we find that the total contribution to the Liouville form
(\ref{theta}) from the ridge transition region $|r|
\le (1+\ridgepar)\rho$ is
\begin{equation}
\int_{-(1+\ridgepar)\rho}^{(1+\ridgepar)\rho} dr \,
( \pi_\Lambda \extder\Lambda + \pi_R \extder R )
=
\left[
\rho
\ln
\left(
\sqrt{2 M_-} + \sqrt{\rho}
\over
\sqrt{2 M_-} - \sqrt{\rho}
\right)
- 2 \sqrt{2M_-\rho}
\right]
\extder \rho
+ o(1)
\ \ .
\label{ridge-trans-cont}
\end{equation}

\section{Shell transition region}
\label{app:shell-trans}

In this appendix we specify the gauge in the shell transition region
$\zr - \shellpar \le r \le \zr$, and evaluate the contribution from this
region to the integral on the right-hand side of (\ref{theta}) in the limit
$\shellpar\to0$.

\subsection{Gauge choice}
\label{subapp:shell-gauge}

To specify the gauge in the region $\zr - \shellpar \le r \le \zr$, we again
consider the classical spacetime of subsection~\ref{subsec:gauge-choice},
and the spacelike hypersurface ${\tilde\Sigma}_0$ in this spacetime. The
part $\zr - \shellpar \le r \le \zr$ of ${\tilde\Sigma}_0$ lies in the
Kruskal spacetime left of the shell.

Let $f:\BbbR\to\BbbR$ be defined by
\begin{equation}
f(x) := \left\{
\begin{array}{l}
x\, e^{-x^2/(1-x^2)} \\ 0 \end{array} \begin{array}{l} , \hspace{2mm} x \in
(0,1) ,\\ , \hspace{2mm} x \notin (0,1) . \end{array}
\right.
\label{f-def}
\end{equation}
We write $f^{(n)}(x) := d^n f(x)/dx^n$. $f$~is continuous everywhere,
and smooth except at $x=0$, with
$f^{(1)}(x) \to 1$ as $x\to 0_+$ and
$f^{(1)}(x) \to 0$ as $x\to 0_-$.
Note that
$f^{(2)}(x) \to 0$ as $x\to 0_\pm$.

For $(1+\ridgepar)\rho < r < \infty$, we choose the gauge
\begin{mathletters}
\label{shellpar-gauge}
\begin{eqnarray}
\Lambda &=& 1
\ \ ,
\label{shellpar-gauge-Lambda}
\\
R &=& r - {\shellpar \sqrt{{\zp}^2 + m^2} \over \zr}
\,
f \! \left({\zr-r \over \shellpar}\right)
\ \ .
\label{shellpar-gauge-R}
\end{eqnarray}
\end{mathletters}%
Outside the shell transition region $\zr -
\shellpar \le r \le \zr$, this clearly agrees with the spatially flat
gauge~(\ref{simpleflatgauge}). To show that the gauge is admissible, we
first note that for $\zr-\shellpar \le r < \zr$, the constraints are solved
by the gravitational momenta given by~(\ref{pi-sol}). At the shell, the
Hamiltonian constraint (\ref{delta-ham}) is identically satisfied. The
momentum constraint (\ref{delta-mom}) at the shell reads,
using~(\ref{pi-lambda-sol}),
\begin{equation}
\zp = \zr \sqrt{{(R'_-)}^2 -1
+ {2M_-/\zr}}
- \sqrt{2M_+\zr}
\ \ ,
\label{gdelta-mom}
\end{equation}
where (\ref{shellpar-gauge-R}) gives
\begin{equation}
R'_- = 1 + {\sqrt{\zp^2+m^2}\over\zr}
\ \ .
\label{gRprimeminus}
\end{equation}
The constraints can therefore be solved both at the shell and away from the
shell, and the gauge is thus admissible. The gauge is smooth everywhere
except at the shell, and at the shell it is consistent with the regularity
assumptions of section~\ref{sec:action}.

\subsection{Liouville form}
\label{subapp:shell-Liouville}

We now evaluate the contribution to the integral in the Liouville form
(\ref{theta}) from $\zr - \shellpar \le r \le \zr$, in the limit
$\shellpar\to0$.

As the gauge (\ref{shellpar-gauge}) is smooth for $(1+\ridgepar)\rho < r <
\infty$ except at the shell, the differentials $\extder
\Lambda(r)$ and $\extder R(r)$ do not contain delta-functions in $r$
except possibly at $r=\zr$. Equation
(\ref{shellpar-gauge-Lambda}) shows that $\extder\Lambda(r)=0$ everywhere.
It is therefore sufficient to consider separately $\pi_R \extder R$ for
$\zr - \shellpar < r < \zr$, and the  delta-function contribution to $\pi_R
\extder R$ at $r=\zr$.

For $\zr - \shellpar < r < \zr$, (\ref{shellpar-gauge-R}) gives
\begin{mathletters}
\begin{eqnarray}
R' &=&
1 + {\sqrt{{\zp}^2+m^2} \over \zr}
\, f^{(1)}
\ \ ,
\label{gRprime}
\\
R'' &=&
- {\sqrt{{\zp}^2+m^2} \over \shellpar \zr}
\, f^{(2)}
\ \ ,
\label{gRprimeprime}
\\
\extder R &=&
(1-R') \extder\zr
+ O(\shellpar)
\ \ ,
\label{gRdot}
\end{eqnarray}
\end{mathletters}%
where the argument of $f$ and its derivatives is $(\zr-r)/\shellpar$.
{}From (\ref{pi-R-sol}) we have
\begin{equation}
\pi_R =
{\zr R''
\over
\sqrt{R'^2-1 + 2M_-/\zr}
}
\ + O(1)
\ \ ,
\label{gpiR}
\end{equation}
where we have used the observations $R''=O(\shellpar^{-1})$
and $R=\zr +O(\shellpar)$. Note that the first term in (\ref{gpiR})
is~$O(\shellpar^{-1})$. We thus obtain, changing the integration variable
from $r$ to $v:=R'$,
\begin{eqnarray}
\int_{\zr-\shellpar}^{\zr} dr \,
\pi_R \extder R
&=&
\zr \extder \zr
\int_{\zr-\shellpar}^{\zr}
{dr \, R'' (1-R')
\over
\sqrt{R'^2-1 + 2M_-/\zr}
}
\ + O(\shellpar)
\nonumber
\\
&=&
\zr \extder \zr
\int_1^{R'_-}
{dv \, (1-v)
\over
\sqrt{v^2-1 + 2M_-/\zr}
}
\ + O(\shellpar)
\nonumber
\\
&=&
\Biggl[
\sqrt{2M_-\zr} - \sqrt{2M_+\zr} - \zp
\nonumber
\\
&&
\quad
+ \zr \ln
\left(
\zr + \zp + \sqrt{{\zp}^2+m^2} + \sqrt{2M_+\zr}
\over
\zr + \sqrt{2M_-\zr}
\right)
\Biggr]
\extder \zr
\ + O(\shellpar)
\ \ ,
\label{shell-trans-cont}
\end{eqnarray}
where we have used (\ref{gRprimeminus}) for~$R'_-$.

What remains is the delta-function in $\extder R$ at $r=\zr$. {}From
(\ref{shellpar-gauge-R}) we have
\begin{equation}
\extder R = - {\shellpar \sqrt{{\zp}^2 + m^2} \, \extder\zr \over \zr}
\,
\delta(r-\zr)
\ \ + \ \hbox{(non-distributional function of $r$)}
\ \ .
\label{delta-in-extR}
\end{equation}
{}From~(\ref{pi-R-sol}), we have
\begin{mathletters}
\begin{eqnarray}
\pi_R^+
&=&
\casehalf \sqrt{2M_+/\zr}
\ \ ,
\\
\pi_R^-
&=&
{
{(\zp + \sqrt{2M_+\zr})}^2 - M_-\zr
\over
\zr (\zp + \sqrt{2M_+\zr})
}
\ \ ,
\end{eqnarray}
\end{mathletters}%
where we have used~(\ref{gRprimeminus}) and the fact that $R''_-=0$.
As $\pi_R$ is not continuous at $r=\zr$, the product $\pi_R \extder R$ is
not defined as a distribution, and the contribution to the Liouville form is
ambiguous. However, as $\pi_R^\pm$ are both~$O(1)$, and the delta-function
in $\extder R$ (\ref{delta-in-extR}) is~$O(\shellpar)$, we argue that the
ambiguous contribution can be taken to vanish in the limit
$\shellpar\to0$. It is seen in the main text that this leads to a reduced
Hamiltonian system that reproduces the correct dynamics.

The ambiguity in $\pi_R \extder R$ appears to have the same origin as the
ambiguity of the equation of motion (\ref{rhatmom-eom}) in the unreduced
formalism: both involve varying the action with respect to~$\zr$. Note
that if the function $f$ had been chosen so that $f^{(2)}(x)\not\to0$ as
$x\to0_+$, $R''_-$~and $\pi_R^-$ would be nonvanishing and proportional
to~$\shellpar^{-1}$, and the above argument for the vanishing of the
ambiguity in the limit $\shellpar\to0$ would not apply.

\end{document}